\documentclass{article}

\usepackage{pifont}

\usepackage[nonatbib, preprint]{neurips_data_2023}

\usepackage[utf8]{inputenc} 
\usepackage[T1]{fontenc}    
\usepackage{hyperref}       
\usepackage{url}            
\usepackage{booktabs}       
\usepackage{amsfonts}       
\usepackage{nicefrac}       
\usepackage{microtype}      
\usepackage{xcolor}         
\usepackage{algorithm}
\usepackage{algorithmicx}
\usepackage{algpseudocode}
\usepackage{enumitem}

\usepackage{array}
\usepackage{cellspace}
\addparagraphcolumntypes{c}
\usepackage{graphicx}
\graphicspath{{./Figures}}

\usepackage{subcaption}
\usepackage{multirow}
\usepackage{subcaption}
\usepackage{amsmath}
\usepackage{caption}
\usepackage[flushleft]{threeparttable}

\usepackage{minitoc}

\usepackage[square,numbers]{natbib}
\title{DORIS-MAE: \\
Scientific Document Retrieval using Multi-level Aspect-based Queries
}

\author{
  Jianyou Wang
  \thanks{Equal Contribution}
  \And
  Kaicheng Wang \footnotemark[1]
  \And
  Xiaoyue Wang
  \And 
  Prudhviraj Naidu
  \And \And
  Leon Bergen \thanks{Equal Contribution}
  \And 
  Ramamohan Paturi \footnotemark[2]
  \And
 \textnormal{Laboratory for Emerging Intelligence}\\
  University of California, San Diego\\
  La Jolla, CA 92093 \\
  \texttt{\{jiw101, kaw036, xiw027, prnaidu, lbergen, rpaturi\}} @ ucsd.edu}

\begin{document}

\maketitle

\begin{abstract}
In scientific research, the ability to effectively retrieve relevant documents based on complex, multifaceted queries is critical. Existing evaluation datasets for this task are limited, primarily due to the high cost and effort required to annotate resources that effectively represent complex queries. To address this, we propose a novel task,  \textbf{S}cientific \textbf{DO}cument \textbf{R}etrieval using \textbf{M}ulti-level \textbf{A}spect-based qu\textbf{E}ries (DORIS-MAE), which is designed to handle the complex nature of user queries in scientific research. We developed a benchmark dataset within the field of computer science, consisting of 100 human-authored complex query cases. For each complex query, we assembled a collection of 100 relevant documents and produced annotated relevance scores for ranking them. Recognizing the significant labor of expert annotation, we also introduce Anno-GPT, a scalable framework for validating the performance of Large Language Models (LLMs) on expert-level dataset annotation tasks. LLM annotation of the DORIS-MAE dataset resulted in a 500x reduction in cost, without compromising quality. Furthermore, due to the multi-tiered structure of these complex queries, the DORIS-MAE dataset can be extended to over 4,000 sub-query test cases without requiring additional annotation. We evaluated 17 recent retrieval methods on DORIS-MAE, observing notable performance drops compared to traditional datasets. This highlights the need for better approaches to handle complex, multifaceted queries in scientific research. Our dataset and codebase are available at \href{https://github.com/Real-Doris-Mae/Doris-Mae-Dataset}{https://github.com/Real-Doris-Mae/Doris-Mae-Dataset}.
  
\end{abstract}
\section{Introduction}
\label{sec: introduction}
\begin{figure}[t]
    \centering
        \includegraphics[width=\linewidth]{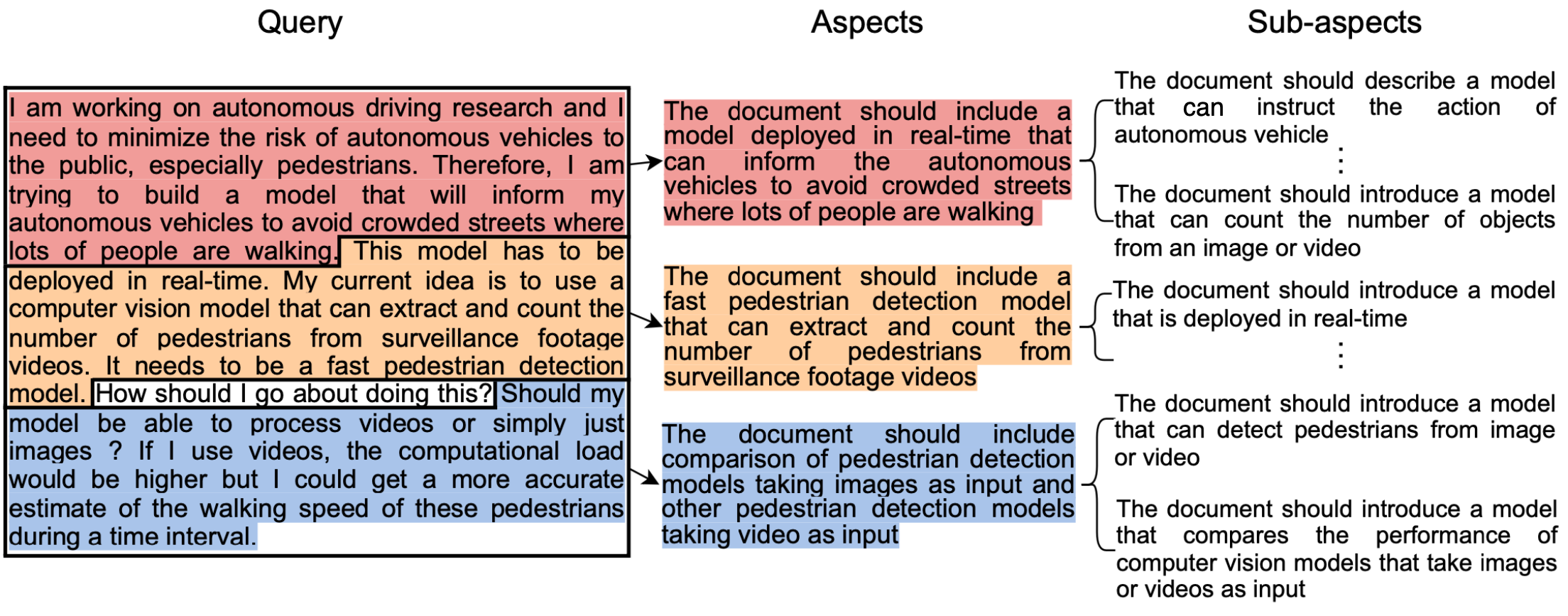}
    \caption{Example from the DORIS-MAE dataset. Each query is broken down into aspects and sub-aspects. Aspects are semantically distinct components of the query, and sub-aspects are minimal requirements that can be extracted from the aspects.}
    \label{fig:query breakdown}
\end{figure}

Scientists often have complex questions that require thorough exploration within various parts of their field (Figure \ref{fig:query breakdown}). Finding relevant scientific literature, one of many challenges in this process, can be especially difficult when dealing with multi-faceted queries. These queries typically encompass numerous interconnected topics and require an information retrieval (IR) system capable of recognizing and responding to this level of complexity.

Information retrieval, especially query-based document retrieval \cite{document_retrieval1, document_retrieval2,passage_retrieval}, is integral to many applications, from search engines \cite{search_engine1,search_engine2, search_engine3} and content recommendations \cite{recommendation0,recommendation1, recommendation2}, to open-domain question answering (QA) \cite{NeuralQA1, BiomedicalQA,open_domain_QA}. A persistent challenge, however, is the low accuracy in processing complex and multi-intent user queries. Despite advanced search engines using semantic understanding and user behavior data in addition to keyword matching \cite{search_engine2}, these systems still fall short when dealing with complex, multi-intent queries. Neural Information Retrieval (NIR) models \cite{rocketqa_v2, ance,simlm, spladev2, colbertv2, ance, ernie} are primarily trained on relatively simple queries \cite{msmarco, qadataset1, qadataset2, qadataset3, qadataset4}. Moreover, \citet{abnirml} creates a variety of diagnostic probes, revealing NIR models' instability when processing out-of-distribution textual inputs, potentially because they do not comprehend the deeper semantics of text. These limitations can lead to inadequate performance with more complex queries. While open-domain QA models like ChatGPT \cite{GPT_few_shot_learner} have shown significant capability, they frequently produce incorrect or even fabricated responses \cite{hallucination1, hallucination2, hallucination3, hallucination4, hallucination5}, and are prohibitively expensive if directly applied to an entire corpus of scientific documents.

To address these challenges, we introduce a novel task, \textbf{S}cientific \textbf{Do}cument \textbf{R}etrieval for \textbf{M}ulti-level \textbf{A}spect-based  qu\textbf{E}ries (DORIS-MAE). DORIS-MAE extends query-based and example-based IR paradigms \cite{qbe1, qbe2}, aiming to give users more control in formulating queries using natural language.
In order to advance research in this area, we present the DORIS-MAE dataset, comprising 100 unique complex queries in the computer science domain, paired with ranked pools of relevant CS article abstracts. Each query is organized into a hierarchical structure of aspects/sub-aspects, which aids annotation.

A distinguishing feature of the DORIS-MAE dataset is its aspect-based hierarchical structure shown in Figure \ref{fig:query breakdown}. This feature aids automation of the annotation process, expands the test case volume without additional annotation, and opens up exploration into the usefulness of aspect information for retrieval methods.

Furthermore, we introduce Anno-GPT, a pipeline for validating Large Language Model (LLM) annotations in a statistically sound manner. Our tests reveal that LLM annotations achieve quality comparable to those produced by human annotators, but with considerable savings in both time and cost. Additionally, the design of our pipeline lends itself to easy adaptation for different domains.

In our experiments, we evaluated 17 IR/NIR (Information Retrieval/Neural Information Retrieval) methods using the DORIS-MAE dataset. The methods have worse performance on DORIS-MAE compared to traditional document retrieval datasets, highlighting the complexity of DORIS-MAE, and the need for more sophisticated retrieval methods. 

Our main contributions are three-fold. First, by formulating the \textbf{DORIS-MAE} task we are shedding a new light on tackling complex, multi-faceted queries during scientific research. Second, we propose \textbf{Anno-GPT}, a procedure for rigorously evaluating the ability of LLMs to replace human experts for challenging annotation tasks. Third, we demonstrate the value of breaking complex queries down to an \textbf{Multi}-level \textbf{Aspect}-based hierarchical structure, for both annotation accuracy and potential improvements in retrieval methods. 

\section{Related Work}
\label{sec: related work}
A range of methods have been developed for document retrieval and re-ranking. Classic retrieval methods like TF-IDF \cite{tfidf} and BM25 \cite{bm25} utilize keyword matching between queries and documents, and fall short when the necessary key phrases are not known to users. To address this, researchers have applied deep learning techniques to develop NIR models, including RocketQA-v2 \cite{rocketqa_v2}, ColBERT-v2 \cite{colbertv2}, SimLM \cite{simlm}, and SPLADE-v2 \cite{spladev2}. These models generate latent vector representations for queries and documents, which are effective for many retrieval tasks, but may not be able to handle textual inputs that are outside of their training distributions \cite{abnirml}.

Other models such as SPECTER \cite{specter} and aspect-based ASPIRE \cite{aspire} focus on calculating document-level similarity. These models are designed to retrieve given an existing paper as input, which makes them less suited for open-ended queries. Other retrieval models, such as text-embedding models (ada-002 by OpenAI \cite{openaiada} and E5-large-v2 \cite{e5largev2}) and sentence-embedding model Sentence-BERT \cite{sentbert}, all face similar limitations as NIR methods in that they struggle to simultaneously represent multiple aspects of a query. Despite these challenges, they tend to perform better on our dataset than specialized NIR models for efficient dense passage retrieval, indicating their potential for complex tasks like DORIS-MAE.

Large-scale traditional IR datasets \cite{msmarco, wikiqa, NQ, colbertv2, beir} primarily contain simple, web-based queries. Models performing well on these datasets often struggle on DORIS-MAE. In contrast, more specialized IR datasets \cite{neves_2019, solvent, csfcube} require human expert annotation, and consequently are more limited in their scale. \citet{csfcube} and \citet{solvent} introduce the concept of "aspect" in document retrieval tasks. These datasets provide pre-defined categories for aspects such as "background" or "method". DORIS-MAE extends this work by allowing for open-ended aspects based on the context of the queries. In Narrative-Driven Recommendation (NDR) research \cite{NDR1,NDR2, NDR3}, user queries are descriptions that capture a range of users' needs. These queries are related to everyday tasks, such as finding restaurants and entertainment.

Progress in LLMs \cite{GPT_few_shot_learner} and prompt-engineering \cite{prompt1, prompt2, prompt3, prompt4, prompt5} has made it feasible to leverage LLMs such as ChatGPT for annotating NLP tasks \cite{AnnoLLM, gpt3_annotator}, even outperforming crowd-workers in some cases \cite{chatgpt_task_solver, chatgpt4_political, chatgpt_outperform_crowdworkers}. However, these tasks do not necessitate domain-specific knowledge, and the annotations produced may not measure up to expert annotations. \citet{LLM_annotation1_faggioli, LLM_annotation2_macavaney} explored the notion of using LLMs to assist human in relevance judgements, and evaluated on TREC-DL datasets \cite{trec-dl2020}, which have single-faceted queries. Our work seeks to extend these efforts to new annotation tasks requiring domain expertise, while introducing a separate pipeline stage and a hypothesis testing stage.

\section{Dataset Description}
\label{sec: dataset description}

The DORIS-MAE task uses a dataset of 100 complex, human-written queries, each containing between 95 to 226 words. Each complex query is broken down into a hierarchy of aspects and sub-aspects, with aspects representing significant semantic components of the original query, typically a sentence or a few sentences. Sub-aspects further decompose an aspect into simpler, verifiable requirements. Both aspects and sub-aspects are generally one sentence long, though their semantic complexity varies. A complex query can have up to 9 aspects, and each aspect can further contain up to 6 sub-aspects. Figure \ref{fig:query breakdown} shows an example.

For each complex query $Q$, we created a pool $\mathcal{P}$ of approximately 100 potentially relevant scientific abstracts to evaluate the re-ranking performance of various retrieval/reranking methods. Within the context of $Q$ and its $\mathcal{P}$, the collection of aspects and sub-aspects is together denoted as (with slight abuse of notation) $Q:= \{a_i\} $. For any aspect or sub-aspect $a_i \in Q$, and any paper abstract $p_j \in \mathcal{P}$, they form a question pair $(a_i,p_j)$. We generated a total of 165,144 question pairs from the 100 queries, together denoted as $\mathcal{D}:= \{(a_i,p_j)\}$. 

To compute the relevance of a paper $p_j$ for a query $Q$, we first break the query down into its aspects. We then compute the relevance score $S(a_i,p_j)$, which measures the relevance of paper $p_j$ for aspect $a_i$. The relevance score $S(p_j|Q)$ for the query $Q$ is the sum of its aspect relevance scores. An average score $\frac{S(p_j|Q)}{|Q|} \geq 1$ indicates $p_j$ is a relevant abstract for $Q$. 
\vspace{-0.5em}
\begin{equation}
\label{eq:full-q}
    S(p_j|Q) := \sum_{a_i \in Q} S(a_i, p_j)
\end{equation}
Thus, for each complex query and candidate pool, we provide a complete ranking for the list of abstracts. Additionally, we can choose a combination of aspects within a complex query $Q$ and concatenate the corresponding sentences to form a sub-query $q \subset Q$, and can calculate the relevance score between a paper abstract $p_j$ and a sub-query $q$.
\vspace{-0.5em}
\begin{equation}
\label{eq:sub-q}
    S(p_j|q) := \sum_{a_i \in q} S(a_i, p_j)
\end{equation}
Hence, for any sub-query $q \subset Q$ and the candidate pool $\mathcal{P}$ of $Q$, we can also provide a complete ground-truth ranking order for $\mathcal{P}$. This allows the dataset to extend to over 4000 sub-query test cases at no additional annotation cost.

Candidate pool abstracts are taken from a database of 360,000 computer science papers between 2011-2021 sourced from arXiv\footnote{https://arxiv.org/}. We complemented each arXiv paper with its corresponding citation information by cross-matching it on Semantic Scholar \cite{semantic_scholar}.

\subsection{Query Formation}
\label{ssec:query formation}

The DORIS-MAE task aims to mirror real-world scenarios where a researcher has an incomplete concept for a research project and needs to explore the breadth of existing research to establish a solid starting point. The 100 complex queries in our dataset simulate this scenario.

Each query is based on one or more existing research papers. We randomly selected 140 papers from AI, NLP, ML, and CV categories on arXiv.
We examined each selected paper's motivation, background, related work, methodology, and experimental results. Using this information, we reverse-engineered a complex query designed to reflect the early thought process of the paper's authors. Overall, DORIS-MAE contains 80 queries derived from single paper abstracts, with 20 queries from each of AI, NLP, ML and CV. We created an additional 20 composite queries, each integrating ideas from 2-3 abstracts. These composite queries are designed to simulate more interdisciplinary and unexplored research ideas than the other 80 queries. 

\subsection{Decomposing Queries to Aspects}
\label{ssec: aspect}

In this section, we discuss our process of decomposing queries into a list of aspects and further breaking down aspects into sub-aspects. The guiding principles for determining aspects are as follows: 
\begin{enumerate}[label=(\roman*)]
    \item Each aspect must correspond to a prominent and semantically meaningful component in the query (refer to Figure \ref{fig:query breakdown} for this correspondence).
    \item Each aspect should contain sufficient context to make sense independently, eliminating potential ambiguities.
    \item Each aspect must be semantically distinct and unrelated to others, ensuring their contents are disjoint and do not overlap.
\end{enumerate}
Some aspects may fall into broader facets such as \textit{background}, \textit{method}, or \textit{result} as defined in \cite{csfcube}. Regarding decomposing aspects to sub-aspects, our criteria are:
\begin{enumerate}[label=(\roman*)]
    \item Each sub-aspect should not contain more information than its parent aspect.
    \item Each sub-aspect should represent a semantic segment of its parent aspect
    \item Different sub-aspects may overlap semantically but each should pose a distinct question.
    \item Each sub-aspect should be as simple as possible.
\end{enumerate}


Given the inherent difficulty in recalling the full details of a query while ranking a large candidate pool, the aspect-based hierarchical structure of our complex queries provides a systematic, efficient, and interpretable approach for annotation. This leads to a more precise ranking of the candidate pool. As each aspect corresponds to a specific part of the original query $Q$, combining several aspects is equivalent to concatenating their corresponding parts in the query, forming semantically coherent sub-queries $q \subset Q$.

\subsection{Candidate Pooling}
To create a candidate pool $\mathcal{P}$ of approximately 100 paper abstracts for each query $Q$, we used a variety of IR and NIR methods, similar to the approach in \cite{csfcube}. More specifically, we utilized popular IR search algorithms TF-IDF \cite{tfidf} and BM25 \cite{bm25} at different granularities (i.e., sentence and paragraph-level for each query) to retrieve around 80 paper abstracts. We then employed OpenAI's text embedding model, ada-002 \cite{openaiada}, to extract 20 more paper abstracts. We also use citation signals from semantic scholar \cite{semantic_scholar} to included any papers that directly cite or are cited by any of the original papers used to create that specific query. Lastly, to prevent bias towards lexicon-matching retrieval methods, we excluded from $\mathcal{P}$ any papers that authors referenced during query $Q$'s formulation. See Appendix \ref{ssec:extending candidate pools} for a sensitivity analysis of our candidate pool construction procedure.

\section{Anno-GPT Framework}
\label{sec: anno-gpt framework}
We propose Anno-GPT, a framework for developing a statistically sound annotation pipeline. We use chatgpt-3.5-turbo-0301 to annotate all 165,144 question pairs $\mathcal{D}:=\{(a_i, p_j)\}$ where $a_i$ is an aspect/sub-aspect of a query, and $p_j$ is an abstract in the query's candidate pool. This strategy minimizes human annotation efforts. The key to this approach lies in breaking down complex queries into simpler aspects and sub-aspects, ensuring an objective and manageable annotation task. Without this structured approach, we found that evaluating an abstract's relevance to a complex query was challenging, due to variability in how partial relevance was assessed. However, the scenario changes significantly when dealing with question pairs $(a_i, p_j)$, which only look at a single aspect or sub-aspect of the query. In such cases, assigning a coarse-scale relevance score between 0-2 becomes feasible, maintaining a reasonable degree of objectivity.

The performance of the annotation pipeline may be influenced by several factors: the procedure for breaking down queries into aspects, the criteria used for scoring query relevance, and the LLM prompt selection \cite{prompt1, prompt2, prompt3, prompt4, prompt5}. In order to avoid overfitting in this pipeline, our methodology comprises two distinct development and testing stages.
The development stage involves optimizing all stages of the pipeline, and using feedback and observed outcomes to iteratively refine this strategy.
The testing stage uses a prespecified hypothesis test. The null hypothesis is that there is no significant difference between the agreement levels of ChatGPT and humans, and those among humans themselves.
After the pipeline has been optimized, we evaluate it on a test set. We compute the difference between ChatGPT-human agreement and human-human agreement. If this difference is sufficiently close to 0, with a small confidence interval, we can use the optimized prompt $\pi$ to annotate the full dataset. 

Below we give a summary of the Anno-GPT framework:
\vspace{-0.5em}
\begin{enumerate}[label=\roman*, itemsep=0pt, parsep=0pt]
    \item Construct aspect-document question pairs, $\mathcal{D} := \{(a_i,p_j)\}$.
    \item Select development set $S\textsubscript{dev} \subset \mathcal{D}$, and use human annotators to score $S\textsubscript{dev}$.
    \item Select test set $S\textsubscript{test} \subset \mathcal{D} \setminus S\textsubscript{dev}$ based on desired power, and use human annotators to score $S\textsubscript{test}$. 
    \item Optimize prompting strategy $\pi$ and other hyperparameters on $S\textsubscript{dev}$. Fix $\pi$.
    \item Evaluate fixed $\pi$ on $S\textsubscript{test}$.
    \item If satisfactory performance is achieved on $S\textsubscript{test}$, proceed to use $\pi$ to annotate the entire dataset $\mathcal{D}$. 
    \item Otherwise, repeat steps iii, iv, v for $S\textsubscript{dev} \leftarrow S\textsubscript{dev} \cup S\textsubscript{test}$ and new $S\textsubscript{test}$. 
\end{enumerate}
\vspace{-0.5em}
It is important to note that automated annotation for DORIS-MAE only used a single development/testing cycle, and therefore did not go into step vii. Therefore, there was no risk of inflated estimates of annotation accuracy due to multiple comparisons \cite{multiple_comparisons, multiple_comparisons_bonferroni, multiple_comparison_sidak}. If multiple cycles are necessary, then the new $S\textsubscript{test}$ must be sufficiently large to avoid these problems.

\subsection{Annotation Guidelines}
Our team of annotators consists of three graduate students in computer science, all with at least two years of research experience in NLP, CV, ML, and AI. Annotators, both human and ChatGPT, are asked to score each question pair using a 3-point grading scale (0-2):
\vspace{-0.5em}
\begin{itemize}
    \item \textbf{Level 0:} The abstract is unrelated or does not provide any help to the key components of the aspect or sub-aspect.
    \item \textbf{Level 1:} The abstract answers some (or all) key components (either explicitly or within one natural inferential step), but at least one key component is not answered explicitly.
    \item \textbf{Level 2:} The abstract directly answers all the key components of the aspect or sub-aspect.
\end{itemize}
\vspace{-0.5em}
We decided to include both direct and indirect coverage under Level 1, acknowledging that distinguishing between these cases could be challenging and potentially subjective. The detailed guidelines for human annotation can be found in Appendix \ref{sec:annotation}.

\subsection{Optimization of Annotation Pipeline}

 For the development stage, two annotators independently annotated a randomly selected set $S\textsubscript{dev}$ of 90 question pairs from the complete set $\mathcal{D}$ of 165,144 pairs. This annotated development set then served as the basis for refining the prompting strategies for ChatGPT. 

The quality of annotations was evaluated using three metrics: macro F1 score, exact accuracy (agreement), and Spearman's rank correlation coefficient (Spearman's $\rho$). 
These metrics measure the agreement level between annotators, and have been used successfully in similar tasks \cite{csfcube}. After satisfactory agreement levels were achieved between ChatGPT and human annotations on the development set, we transitioned to the hypothesis testing stage, with all three annotators involved. In this stage, we employed the fixed finalized prompting strategy, ensuring no overfitting or leakage from the test set.

Our prompting strategy development involved experimenting with recent methodologies such as few-shot in-context learning (ICL) \cite{icl1, GPT_few_shot_learner, icl3, icl4, icl5} and chain of thought (CoT) \cite{cot1, cot2, cot3, cot4}. We found that the CoT approach offered the most robust and optimal results for the task of annotating question pairs $(a_i,d_j) \in \mathcal{D}$. A comprehensive description of the prompt engineering process can be found in Appendix \ref{sec:annotation}.

\subsection{Annotation Evaluations}

Hypothesis testing was conducted using a sample of 250 question pairs $S\textsubscript{test}$ from $\mathcal{D}$, distinct from the development set. The selected pairs were independently annotated by three human annotators. We used bootstrapping to estimate the 95\% confidence intervals for the macro F1 score, accuracy (agreement), and Spearman's $\rho$. Though the sampling temperature is fixed at zero, recognizing small randomness introduced by GPU non-determinism \cite{gpu}, the ChatGPT annotations are run twice and the pairwise comparisons with human annotators are averaged across these runs.


In addition to these prespecified analyses, we conducted post-hoc analyses using an adjudication procedure to create a more stable set of human annotations \cite{csfcube}. We use majority voting \cite{majority_vote} between the three human annotators to decide the adjudicated annotation. 

The results presented in Table \ref{tab:hypothesis-analysis} show that the rate of ChatGPT-human agreement is within range of human-human agreement. Specifically, ChatGPT's performance is comparable to that of human annotators as measured by F1 and exact agreement (accuracy). The average agreement level for ChatGPT is numerically lower than average human agreement level as measured by Spearman's $\rho$.
We note that the lowest Spearman's $\rho$ among two humans is $46.51\%$, which is comparable to the average ChatGPT/human's $\rho= 46.61\%$, suggesting ChatGPT's performance is still within the range of human-level agreement. All $p$-values are larger than the $\alpha=0.05$ criterion. 

In a post-hoc analysis, we found that comparing ChatGPT to adjudicated human annotations numerically increased the rate of agreement. This provides qualitative evidence for ChatGPT's performance relative to that of human experts. 
We further analyzed instances where ChatGPT's annotations diverged from those of humans. Interestingly, the nature of these discrepancies was similar to those found between humans, with differences largely revolving around the interpretation of key components in aspect/sub-aspects. For example, ChatGPT occasionally differed from human annotators in determining the importance of a given component. Detailed examples of ChatGPT's reasoning and a comprehensive error analysis can be found in Appendix \ref{sec:annotation}.



\begin{table}[ht]
\setlength{\tabcolsep}{3pt} 
\centering
\caption{Annotation agreement between humans and ChatGPT. H is human, G is ChatGPT, A is Adjudication, CI is 95\% confidence interval. $p$-values correspond to the null hypothesis that there is no difference between avg. H\&H and avg. H\&G. Higher $p$-values indicate less evidence of a difference between ChatGPT and humans. \\ \quad}
\label{tab:hypothesis-analysis}
\begin{tabular}{l|c|cc|cc|c}
\toprule
\textbf{Metrics} &\textbf{G\&A} & \textbf{avg. H\&H} & \textbf{avg. G\&H} & \textbf{H\&H CI} & \textbf{G\&H CI} & \textbf{$p$-value} \\
\midrule
F1-score (macro) & 64.17 &  58.33 & 57.46 & (52.33, 63.46) & (50.93, 62.79) &  0.74\\
Accuracy & 67.40 &  64.13 & 62.07 & (59.73, 68.80) & (57.67, 66.13) &  0.41\\
Spearman’s $\rho$ & 52.63 & 54.31  & 46.61 & (46.87, 61.56) & (38.67, 54.41) & 0.07\\
\bottomrule
\end{tabular}
\end{table}

\subsection{Scalability of Annotations}
\label{ssec:scalability of anno-gpt}
The hypothesis testing results support the use of ChatGPT for annotation. At deployment, the pipeline annotated all 165,144 aspect-paper pairs within a span of 24 hours, at a cost under \$150. By contrast, human experts typically require approximately 4 minutes per question pair, resulting in an estimated 11,146 hours to annotate the entire dataset. The deployment resulted in a time reduction by a factor of 500, and a cost reduction by a factor of 1,000, without sacrificing annotation quality. Upon completion of the annotation process, we utilized Equations \ref{eq:full-q} and \ref{eq:sub-q} to compile the results and compute the final rankings for both full-query and sub-query test cases. Anno-GPT could potentially utilize any high-performance LLM to replace ChatGPT and can be adapted for other expert-level tasks, given the availability of a small set of domain expert annotations for validation.



\section{Retrieval Results}
\label{sec: experiment results}

This section presents the results of testing 17 models discussed in Section \ref{sec: related work} on the DORIS-MAE dataset. When available, we trained the model on our CS corpus and denoted the best version as trained in domain (ID), see full training details in Appendix \ref{ssec:retrain details}. To contextualize their performance on DORIS-MAE, we compare the results with these models' previously reported performances on various IR datasets, including MS MARCO \cite{msmarco}, LoTTE \cite{colbertv2}, NQ \cite{NQ}, and Wiki-QA \cite{wikiqa}.

\subsection{DORIS-MAE Benchmarking Results}

In our benchmarking process for DORIS-MAE, we use complex queries as inputs to these models. We employ a variety of metrics common in the IR/NIR literature for the evaluation, including R@5, R@20, R-Precision (RP), $\textrm{NDCG}^{\textrm{exp}}_{10\%}$, MRR@10, and MAP. For fairness, we adopted an alternative approach for models like RocketQA-v2 \cite{rocketqa_v2} and ColBERT-v2 \cite{colbertv2} that were not designed to handle long queries. For these cases, we allow the models to process the input as either a single text string or sentence-by-sentence, and report the maximum performance achieved. Uniquely among the models that we consider, the ASPIRE models (TSApire/OTAspire) are designed to handle multi-aspect queries. For brevity, we only report the higher number among these two options for models in Table \ref{tab:query-level ranking}. For more detailed results, refer to Appendix \ref{sec:experiment results}. 

\begin{table}[ht]
\setlength{\tabcolsep}{3pt} 
\centering
\caption{Query level performance on full DORIS-MAE. Standard errors are estimated by bootstrapping. ID means a model is trained in domain.}
\vspace{1em}
\footnotesize 
\renewcommand{\arraystretch}{1.0} 
\begin{tabular}{lcccccc}
\toprule
Method & R@5 & R@20 & RP & $\textrm{NDCG}^{\textrm{exp}}_{{10\%}}$ & MRR@10 & MAP\\
\midrule
Random & 4.41 & 18.48 & 16.29 & 7.31 & 3.59 & 19.63\\
\midrule
E5-L-v2\cite{e5largev2} & \textbf{16.51} \scriptsize{$\pm$ 2.05} & 43.77 \scriptsize{$\pm$ 2.14} & \textbf{37.46} \scriptsize{$\pm$ 2.44} & 25.90 \scriptsize{$\pm$ 2.15} & 14.85 \scriptsize{$\pm$ 2.73} & \textbf{40.49} \scriptsize{$\pm$ 2.32} \\
RocketQA-v2\cite{rocketqa_v2} & 15.63 \scriptsize{$\pm$ 1.88} & \textbf{45.41} \scriptsize{$\pm$ 2.43} & 34.36 \scriptsize{$\pm$ 2.32} & \textbf{30.30} \scriptsize{$\pm$ 2.26} & \textbf{20.87} \scriptsize{$\pm$ 3.12} & 40.18 \scriptsize{$\pm$ 2.23} \\
ada-002\cite{openaiada} & 15.38 \scriptsize{$\pm$ 1.95} & 42.84 \scriptsize{$\pm$ 2.53} & 35.81 \scriptsize{$\pm$ 2.67} & 27.46 \scriptsize{$\pm$ 2.48} & 19.88 \scriptsize{$\pm$ 3.21} & 40.37 \scriptsize{$\pm$ 2.55}\\
SimCSE\cite{simcse} & 14.90 \scriptsize{$\pm$ 1.89} & 42.62 \scriptsize{$\pm$ 2.40} & 35.27 \scriptsize{$\pm$ 2.34}  & 26.88 \scriptsize{$\pm$ 2.36} & 21.19 \scriptsize{$\pm$ 3.47} & 39.02 \scriptsize{$\pm$ 2.35}\\
SPLADE-v2\cite{simcse} & 14.78 \scriptsize{$\pm$ 1.89} & 40.14 \scriptsize{$\pm$ 2.33} &31.65 \scriptsize{$\pm$ 2.38}  & 26.08 \scriptsize{$\pm$ 2.00} & 17.82 \scriptsize{$\pm$  2.99} & 37.23 \scriptsize{$\pm$ 2.26}\\
SPECTER-v2\cite{specter} & 14.50 \scriptsize{$\pm$ 2.15} & 43.36 \scriptsize{$\pm$ 2.50} & 33.41 \scriptsize{$\pm$ 2.33} & 25.65 \scriptsize{$\pm$ 2.23} & 17.19 \scriptsize{$\pm$ 2.96} & 37.12 \scriptsize{$\pm$ 2.10} \\
SPECTER\textsubscript{ID} & 13.32 \scriptsize{$\pm$ 1.76} & 42.52 \scriptsize{$\pm$ 2.37} & 31.55 \scriptsize{$\pm$ 2.28} & 21.27 \scriptsize{$\pm$ 2.03} & 14.48\scriptsize{$\pm$ 2.78} & 36.02 \scriptsize{$\pm$ 2.19}\\
TSAspire\cite{aspire} & 14.26 \scriptsize{$\pm$ 1.80} & 41.25 \scriptsize{$\pm$ 2.40} & 33.81 \scriptsize{$\pm$ 2.47} & 26.63 \scriptsize{$\pm$ 2.05} & 15.59 \scriptsize{$\pm$ 2.59} & 37.00 \scriptsize{$\pm$ 2.29} \\
SentBERT\cite{sentbert} & 14.09 \scriptsize{$\pm$ 1.88} & 44.69 \scriptsize{$\pm$ 2.47} & 33.79 \scriptsize{$\pm$ 2.41} & 21.88 \scriptsize{$\pm$ 2.07} & 13.23 \scriptsize{$\pm$ 2.69} & 37.75 \scriptsize{$\pm$ 2.28} \\
OTAspire\cite{aspire} & 13.34 \scriptsize{$\pm$ 1.56} & 42.25 \scriptsize{$\pm$ 2.53} & 33.63 \scriptsize{$\pm$ 2.38} & 25.52 \scriptsize{$\pm$ 2.29} & 14.18 \scriptsize{$\pm$ 2.66} & 36.70 \scriptsize{$\pm$ 2.22} \\
ANCE\textsubscript{FirstP}\cite{ance} & 13.21 \scriptsize{$\pm$ 2.02} & 34.54 \scriptsize{$\pm$ 2.20} & 30.51 \scriptsize{$\pm$ 2.50} & 20.30 \scriptsize{$\pm$ 2.02} & 13.87 \scriptsize{$\pm$ 2.64} & 34.53 \scriptsize{$\pm$ 2.35} \\
SPLADE-v2\cite{spladev2} & 11.80 \scriptsize{$\pm$ 1.86} & 36.59 \scriptsize{$\pm$ 2.12} & 29.90 \scriptsize{$\pm$ 2.20} & 21.35 \scriptsize{$\pm$ 2.12} & 14.30 \scriptsize{$\pm$ 2.77} & 33.98 \scriptsize{$\pm$ 2.23} \\
LLAMA\cite{llama} & 12.74 \scriptsize{$\pm$ 1.82} & 34.51 \scriptsize{$\pm$ 2.36} & 28.33 \scriptsize{$\pm$ 2.14} & 16.65 \scriptsize{$\pm$ 1.68} & 11.78 \scriptsize{$\pm$ 2.45} & 31.29 \scriptsize{$\pm$ 1.99} \\
SimLM\cite{simlm} & 12.68 \scriptsize{$\pm$ 1.77} & 35.67 \scriptsize{$\pm$ 2.49} & 28.90 \scriptsize{$\pm$ 2.42} & 18.91 \scriptsize{$\pm$ 1.86} & 11.29 \scriptsize{$\pm$ 2.44} & 33.06 \scriptsize{$\pm$ 2.34} \\
BM25\cite{bm25} & 8.47 \scriptsize{$\pm$ 1.80} & 30.50 \scriptsize{$\pm$ 2.38} & 21.94 \scriptsize{$\pm$ 2.03} & 13.23 \scriptsize{$\pm$ 1.97} & 9.19 \scriptsize{$\pm$ 2.46} & 25.99 \scriptsize{$\pm$ 1.68} \\
ColBERT-v2\cite{colbertv2} & 8.45 \scriptsize{$\pm$ 1.46} & 27.86 \scriptsize{$\pm$ 2.29} & 22.33 \scriptsize{$\pm$ 2.01} & 12.57 \scriptsize{$\pm$ 1.71} & 6.69 \scriptsize{$\pm$ 2.15} & 25.80 \scriptsize{$\pm$ 1.83} \\
TF-IDF\cite{tfidf}& 10.71 \scriptsize{$\pm$ 1.48} & 29.22 \scriptsize{$\pm$ 2.25} & 24.79 \scriptsize{$\pm$ 2.06} & 18.25 \scriptsize{$\pm$ 2.01} & 12.41 \scriptsize{$\pm$ 2.53} & 28.77 \scriptsize{$\pm$ 1.81}\\
ERNIE\cite{ernie} & 6.49 \scriptsize{$\pm$ 0.94} & 22.58 \scriptsize{$\pm$ 1.72} & 20.18 \scriptsize{$\pm$ 1.82} & 9.66 \scriptsize{$\pm$ 1.18} & 3.77 \scriptsize{$\pm$ 1.06} & 22.71 \scriptsize{$\pm$ 1.65} \\
SciBERT\cite{scibert} & 5.13 \scriptsize{$\pm$ 1.25} & 17.99 \scriptsize{$\pm$ 1.69} & 17.13 \scriptsize{$\pm$ 1.88} & 7.50 \scriptsize{$\pm$ 1.34} & 3.41 \scriptsize{$\pm$ 1.57} & 20.34 \scriptsize{$\pm$ 1.64} \\
\bottomrule
\end{tabular}
\label{tab:query-level ranking}
\hfill
\end{table}

To better interpret the results, we compare against a random ranking baseline. 
In general, the models show consistent behavior, with larger and more general-purpose models (like E5-Large-V2, RocketQA-v2, ada-002, Specter-v2) faring better than the smaller and more specialized ones (like SciBERT \cite{scibert}, ColBERT-v2, BM25, TF-IDF). Though the Aspire models were designed for multi-aspect queries, they do not have strong performance on the complex queries in DORIS-MAE.

When we compare the DORIS-MAE performance of these models with their reported results on traditional retrieval datasets MS MARCO (in Table \ref{tab: MS MARCO}) and NQ (in Table \ref{tab: NQ}), we observe a significant reduction in their performances on DORIS-MAE. We choose metrics for comparison based on what is available in previously published work. The results highlight the challenges posed by DORIS-MAE and suggest gaps in the ability of existing methods to generalize well to complex query retrieval.

Finally, in Table \ref{tab: comparison with CSFCube}, we make a comparison with the model performances on specialized retrieval datasets such as CSFCube \cite{csfcube}, RELISH \cite{relish}, and TRECCOVID \cite{treccovid}. The comparison reveals a consistent level of difficulty between DORIS-MAE and these completely human-annotated datasets, indicating that DORIS-MAE presents a similarly challenging retrieval task.
\vspace{-1em}
\begin{table}[ht]
  \begin{minipage}{0.45\textwidth}
    \centering
    \footnotesize
    \caption{Comparison with MS MARCO. Stats collected from \cite{ance,colbertv2,rocketqa_v2,simlm,spladev2}.\\ \quad }
    \begin{tabular}{lc|c}
      \toprule
      & \multicolumn{1}{c|}{MS MARCO} & \multicolumn{1}{c}{DORIS} \\
      Ranking Method  & MRR@10 & MRR@10 \\
      \midrule
      ANCE & 33.0 & 13.87 \\
      ColBERT-v2 & 39.7 & 6.69  \\
      RocketQA-v2 & 41.9 & 20.87 \\
      SimLM & 41.1 & 11.29  \\
      SPLADE-v2 & 36.8 & 14.30  \\
      \bottomrule
    \end{tabular}
    \label{tab: MS MARCO}
  \end{minipage}%
  \hfill
  \begin{minipage}{0.45\textwidth}
    \centering
    \footnotesize
    \caption{Comparison with NQ. Stats collected from \cite{colbertv2}. \\ \quad }
    \begin{tabular}{lc|c}
      \toprule
      & \multicolumn{1}{c|}{NQ} & \multicolumn{1}{c}{DORIS} \\
      Ranking Method  & R@20  & R@20  \\
      \midrule
      ANCE\textsubscript{FirstP}& 81.9 &  34.54 \\
      BM25 & 59.1 &  30.50  \\
      RocketQA-v2 & 83.7 &  45.41 \\
      SimLM &  85.2 & 35.67 \\
      \bottomrule
    \end{tabular}
    \label{tab: NQ}
  \end{minipage}
  \end{table}

\begin{table}[ht]
\caption{Comparison with CSFCube, TRECCOVID, and RELISH. Results are from \cite{csfcube, aspire}.\\ \quad}
\footnotesize
\centering
\begin{tabular}{lccc|c|c|ccc}
\toprule
& \multicolumn{3}{c|}{CSFCube}&\multicolumn{1}{c|}{TRECCOVID}&\multicolumn{1}{c|}{RELISH} & \multicolumn{3}{c}{DORIS-MAE} \\
 Ranking Method  & RP & MAP & R@20 & MAP & MAP & RP & MAP & R@20 \\
\midrule
TSAspire & - & 40.26 & - & 26.24 & 61.29 & 33.81 & 37.00 & 41.25 \\
OTAspire & - & 40.79 & - & 30.92 & 62.57 & 33.63 & 36.70 & 42.25 \\
Specter-v2 & 18.32 & - & 52.12 & 28.24 & 60.62 & 33.41 & 37.12 & 43.36\\
TF-IDF & 14.59 & - & 39.69 & - & - &   24.79 & 28.77 & 29.22 \\
BM25 & 13.50 & - & 42.73 & - & - &  21.94 & 25.99 & 30.50 \\
\bottomrule
\end{tabular}
\label{tab: comparison with CSFCube}
\end{table}


\subsection{Additional Experiments}
\label{sec: additional_experiments}
Up until now, the hierarchical aspect-based structure that Anno-GPT utilizes has been hidden from all the evaluated models because of inability of existing methods to break down a complex query automatically. Even though these structures are not readily available for real-life retrieval methods, they may still hold value once this query decomposition process can be fully automated. To explore this potential, we conduct an experiment where instead of using the original query, each retrieval method had access to a concatenated string of all aspects within a query, excluding sub-aspects. The results, as illustrated in Table \ref{tab:ranking based on aspect}, show that Sentence-BERT performs best on four metrics: R@5/20, RP, and MAP. These findings suggest that the use of aspect information could be potentially beneficial to guide retrieval methods.

The hierarchical structure of our dataset can be used to create additional, less complex tasks involving only parts of the query. For instance, by pulling out parts of the query corresponding to 2 Aspects, we are able to generate over 1000 test cases. For this task, we found a significant increase in the number of relevant abstracts. After re-evaluating all models for the sub-query DORIS-MAE, we observed model performance consistent with those seen in previous benchmarks, as indicated in Table \ref{tab:ranking-subquery}. Comparing Table \ref{tab:ranking-subquery} with Table \ref{tab:query-level ranking}, we observed noticeably higher numbers for metrics such as RP, $\textrm{NDCG}^{\textrm{exp}}_{{10\%}}$, MRR and MAP, which are indicators of better model performances on this sub-query task. Since models remain unchanged, this suggest the sub-query task is less challenging than full-query task. This is intuitive since sub-queries are less complex and contain fewer aspects. Overall, the creation of these sub-query tasks underscores the adaptability of our dataset, which could accommodate a range of task complexities under the setting of DORIS-MAE.

\begin{table}[ht]
\setlength{\tabcolsep}{3pt} 
\centering
\caption{Ranking performance given model access to aspects. Full table is in Appendix \ref{sec:experiment results}}
\vspace{1em}
\footnotesize 
\renewcommand{\arraystretch}{1.0} 
\begin{tabular}{lcccccc}
\toprule
Method & R@5 & R@20 & RP & $\textrm{NDCG}^{\textrm{exp}}_{{ 10\%}}$ & MRR@10 & MAP\\
\midrule
ada-002 & 14.09 & 42.23 & 33.56 & 26.54 & \textbf{20.20} & 37.62 \\
SentBERT& \textbf{17.73} & \textbf{45.34} & \textbf{35.67} & 25.00 & 15.52 & \textbf{39.87} \\
RocketQA-v2 & 13.83 & 43.81 & 32.59 & \textbf{27.45} & 16.08 & 37.90 \\
\midrule
\end{tabular}
\label{tab:ranking based on aspect}
\end{table}
\vspace{-2em}
\begin{table}[ht]
\setlength{\tabcolsep}{3pt} 
\centering
\caption{Ranking performance on sub-query (2 Aspects) task. Full table is in Appendix \ref{sec:experiment results}}
\vspace{1em}
\footnotesize 
\renewcommand{\arraystretch}{1.0} 
\begin{tabular}{lcccccc}
\toprule
Method & R@5 & R@20 & RP & $\textrm{NDCG}^{\textrm{exp}}_{{ 10\%}}$ & MRR@10 & MAP\\
\midrule
ada-002 & \textbf{13.49} & \textbf{40.24} & \textbf{47.35} & \textbf{39.00} & \textbf{24.33} & \textbf{51.67} \\
SentBERT & 12.15 & 36.71 & 45.08 & 34.78 & 20.71 & 48.96 \\
RocketQA-v2 & 12.79 & 39.19 & 46.47 & 38.78 & 23.72 & 50.81 \\
\midrule
\end{tabular}
\label{tab:ranking-subquery}
\end{table}

\vspace{1em}
\subsection{Supervised Learning on DORIS-MAE}
To assess the utility of our dataset for training IR models, we conducted an experiment where we allocated 40 queries for training and the remaining 60 for testing. Using supervised contrastive learning (SCL), we optimized a margin-based triplet loss as presented in Equation \ref{eq:triplet-loss}. Each triplet, represented as $(A,P,N)$, consisted of a query and two abstracts. The higher-ranked abstract in the training data served as the positive instance (i.e., $P$), with the other functioning as the negative instance (i.e., $N$). Given each query's candidate pool size of $\geq 100$, we derived multiple triplets. This process yielded 3,000 triplets from the designated 40 training queries. Subsequently, we fine-tuned an E5-L-v2 model, a RoBERTa-based text embedding variant with 355 million parameters, over a single epoch.  
\begin{equation}
\label{eq:triplet-loss}
    L(A,P,N) = \max \Bigg(\frac{<A,N>}{||A||\cdot||N||}-\frac{<A,P>}{||A||\cdot||P||}+m, 0\Bigg) \quad A,P, N \in \mathbb{R}^n, m>0
\end{equation}
When evaluating the model on the 60 test queries, we noted a marked improvement across all metrics with the SCL-trained model in contrast to the pre-trained baseline. This comparison can be found in Table \ref{tab:contrastive learning}. These positive outcomes, achieved using supervised contrastive learning on DORIS-MAE, underscore the utility of our train/test split for model fine-tuning.

Note that Table \ref{tab:query-level ranking} reports models performance on the full DORIS-MAE dataset. We also report models performance on our proposed test set of 60 queries in Appendix \ref{ssec:test set}. 

\begin{table}[ht]
\setlength{\tabcolsep}{3pt} 
\centering
\caption{Comparison of SCL vs pretrained. Standard errors are estimated by bootstrapping.}
\vspace{1em}
\footnotesize 
\renewcommand{\arraystretch}{1.0} 
\begin{tabular}{lcccccc}
\toprule
Method & R@5 & R@20 & RP & $\textrm{NDCG}^{\textrm{exp}}_{{10\%}}$ & MRR@10 & MAP\\
\midrule
SCL-trained E5-v2 & 19.57 \scriptsize{$\pm$ 2.33} & 52.45 \scriptsize{$\pm$ 3.17} & 44.47 \scriptsize{$\pm$ 3.11} & 34.67 \scriptsize{$\pm$ 3.17} & 23.16 \scriptsize{$\pm$ 4.28} & 49.15 \scriptsize{$\pm$ 3.14} \\


pretrained E5-v2 & 14.70 \scriptsize{$\pm$ 1.72} & 42.38 \scriptsize{$\pm$ 2.59} & 38.24 \scriptsize{$\pm$ 2.94} & 26.31 \scriptsize{$\pm$ 2.94} & 14.53 \scriptsize{$\pm$ 3.69} & 40.62 \scriptsize{$\pm$ 2.85}\\
\bottomrule
\end{tabular}
\label{tab:contrastive learning}
\hfill
\end{table}

\vspace{-1em}

\section{Conclusion and Future Work}
\label{sec:conclusion and future work}

This paper introduces a novel task, \textbf{S}cientific \textbf{DO}cument \textbf{R}etrieval using \textbf{M}ulti-level \textbf{A}spect-based qu\textbf{E}ries (DORIS-MAE), aimed at modeling the process of information retrieval in the context of scientific research. We also present a dataset for DORIS-MAE generated using the Anno-GPT framework. 

The results show room for improvement in the performance of current retrieval methods when dealing with DORIS-MAE. Future studies may explore modifications to model architectures and training procedures to better address complex, multifaceted queries.


An understanding of how noise in aspect annotation affects the overall task is an interesting point for future investigation, as it can shed light on how errors in the automated annotations may affect the final candidate pool ranking in DORIS-MAE. 


The hierarchical structure of complex queries, as exemplified in DORIS-MAE, is an area that warrants further attention. Future work might include the development of more sophisticated automated query breakdown methods, potentially drawing from advances in question decomposition \cite{breakdown3, breakdown4}, sequence-to-sequence modelling  \cite{seq2seq1, seq2seq2} and semantic parsing  \cite{semantic_parsing1, semantic_parsing2, semantic_parsing3}.

\textbf{Limitations}: DORIS-MAE currently contains queries and abstracts from the computer science domain. Consequently, models trained on this dataset may not generalize well to other disciplines. An extended, multi-domain version of DORIS-MAE is a logical direction for future work. The task of determining aspect relevance is challenging due to the complexity of the abstracts. Improved annotation guidelines and training (for both humans and models) may address this challenge. Finally, while we harnessed LLMs to streamline the annotation, the generation of queries and their aspect decomposition remains manual. We found the development of a reliable, automated query generation pipeline difficult, but anticipate that advances in LLMs might bridge this gap in the near future.

\clearpage
\bibliography{reference}

\newpage
\appendix

\textbf{Appendix: Table of Contents}

\begin{enumerate}[label=\Alph*.]
    \item Dataset\dotfill \pageref{sec:dataset}
    \begin{enumerate}[label=\arabic*.]
        \item Dataset License and Code License\dotfill \pageref{ssec:dataset license}
        \item Dataset Hosting, Accessibility and Maintenance\dotfill \pageref{ssec:dataset hosting, accessibility and maintenance}
        \item Dataset Overview\dotfill \pageref{ssec:dataset_overview}
        \item DORIS-MAE Structure\dotfill \pageref{ssec:doris-mae structure}
        \item Statistics of Dataset\dotfill \pageref{ssec:statistics of dataset}
    \end{enumerate}
    \item Annotation\dotfill \pageref{sec:annotation}
    \begin{enumerate}[label=\arabic*.]
        \item Annotation Guidelines\dotfill \pageref{ssec:annotation guidelines}
        \item Examples for Annotations\dotfill \pageref{ssec:examples for annotations}
        \item Prompt Engineering\dotfill \pageref{ssec:prompt engineering}
        \item Error Analysis for ChatGPT's Annotation\dotfill \pageref{ssec:error analysis for chatgpt's annotation}
        \item Annotation Scalability\dotfill \pageref{ssec:annotation scalability}
    \end{enumerate}
    \item Experiment Results\dotfill \pageref{sec:experiment results}
    \begin{enumerate}[label=\arabic*.]
        \item Metrics\dotfill \pageref{ssec:metric introduction}
        \item Query Level Benchmark Results\dotfill \pageref{ssec:query level benchmark results}
        \item Using Aspects for Retrieval\dotfill \pageref{ssec:Aspects are visible}
        \item Models Performance on DORIS-MAE Test Set\dotfill \pageref{ssec:test set}
        \item Two-Aspect Sub-Query Level Benchmark Results\dotfill \pageref{ssec:sub-query level benchmark results}
        \item Candidate Pool Sensitivity Analyses\dotfill \pageref{ssec:extending candidate pools}
    \end{enumerate}
    \item Experiment Setup\dotfill \pageref{sec:experiment setup}
    \begin{enumerate}[label=\arabic*.]
        \item Retraining Details\dotfill \pageref{ssec:retrain details}
        \item Model Specifications\dotfill \pageref{ssec: model specifications}
    \end{enumerate}
    \item Author Statement\dotfill \pageref{sec:author statement}
    \item Datasheet for DORIS-MAE\dotfill\pageref{sec:Datasheet for DORIS-MAE}
    
\end{enumerate}

\clearpage
\section{Dataset}
\label{sec:dataset}

\subsection{Dataset License and Code License}
\label{ssec:dataset license}
The DORIS-MAE dataset is made available under the Creative Commons Attribution-NonCommercial (CC-BY-NC) license. This license allows users to share and adapt the dataset under the condition that they provide appropriate credit to the original creators and do not use it for commercial purposes. A copy of the full license can be found at \href{https://github.com/Real-Doris-Mae/Doris-Mae-Dataset/blob/main/dataset_license.md}{https://github.com/Real-Doris-Mae/Doris-Mae-Dataset/blob/main/dataset\_license.md}.

The code used in this paper is released under the MIT License. The MIT License is a permissive open-source license that allows for the free use, modification, and distribution of the code, as long as the original license is included with any derivative work. A copy of the full license can be found at
\href{https://github.com/Real-Doris-Mae/Doris-Mae-Dataset/blob/main/code_license.md}{https://github.com/Real-Doris-Mae/Doris-Mae-Dataset/blob/main/code\_license.md}.

\subsection{Dataset Hosting, Accessibility and Maintenance}
\label{ssec:dataset hosting, accessibility and maintenance}

The DORIS-MAE dataset with its meta-data is released and can be accessed freely at (\href{https://doi.org/10.5281/zenodo.8299749}{\textcolor{black}{https://doi.org/10.5281/zenodo.8299749}}) under the CC-BY-NC license. We commit to regularly maintain the dataset and codebase by incorporating user feedback. We will potentially introduce more features as part of future work in the next version of DORIS-MAE. We confirm that the current version of DORIS-MAE would always remain accessible at the same link.

\subsection{Dataset Overview}
\label{ssec:dataset_overview}

The DORIS-MAE dataset is comprised of four main sub-datasets, each serving distinct purposes.

The \textbf{Query} dataset contains \textcolor{black}{100} human-crafted complex queries spanning across five categories: ML, NLP, CV, AI, and Composite. Each category has \textcolor{black}{20} associated queries. Queries are broken down into aspects (ranging from 3 to 9 per query) and sub-aspects (from 0 to 6 per aspect, with 0 signifying no further breakdown required). For each query, a corresponding candidate pool of relevant paper abstracts, ranging from 99 to \textcolor{black}{138}, is provided.

The \textbf{Corpus} dataset is composed of 363,133 abstracts from computer science papers, published between 2011-2021, and sourced from arXiv. Each entry includes title, original abstract, URL, primary and secondary categories, as well as citation information retrieved from Semantic Scholar. A masked version of each abstract is also provided, facilitating the automated creation of queries.

The \textbf{Annotation} dataset includes generated annotations for all \textcolor{black}{165,144} question pairs, each comprising an aspect/sub-aspect and a corresponding paper abstract from the query's candidate pool. It includes the original text generated by ChatGPT explaining its decision-making process, along with a three-level relevance score (e.g., 0,1,2) representing ChatGPT's final decision.

Finally, the \textbf{Test Set} dataset contains human annotations for a random selection of 250 question pairs used in hypothesis testing. It includes each of the three human annotators' final decisions, recorded as a three-level relevance score (e.g., 0,1,2). Note, the development set contains 90 annotated question pairs. Since the annotations in the development set come from only 2 annotators and since the development set is strictly not a part of the hypothesis testing stage, we do not include the development set in the officially released DORIS-MAE dataset. Instead, we provide the development set as part of the codebase on \href{https://github.com/Real-Doris-Mae/Doris-Mae-Dataset/}{https://github.com/Real-Doris-Mae/Doris-Mae-Dataset/}.

It is important to note that each of these datasets serves a specific function in our overall process, contributing to the optimization, validation, and benchmarking of our proposed approach.

\subsection{DORIS-MAE Structure}
\label{ssec:doris-mae structure}

\begin{enumerate}[label=-]
    \item \texttt{aspect2aspect\_id}: A dictionary that maps aspects/sub-aspects to their aspect\_id.
    \item \texttt{aspect\_id2aspect}: The inverse mapping of \texttt{aspect2aspect\_id}. 
    \item \textbf{Query}: contains 50 complex queries.
    \begin{enumerate}[label={\textbullet}]
        \item Query 1:
        \begin{enumerate}[label=-]
        \item \texttt{query\_text}: The text of the query.
        \item \texttt{query\_type}: One of the [\texttt{ML, NLP, CV, AI, Composite}]. 
        \item \textcolor{black}{\texttt{idea\_from}: The id of the referenced abstract when creating the query}
        \item \texttt{candidate\_pool}: The list of abstract\_ids in the candidate pool of Query 1.
        \item \texttt{sent2aspect\_id}: A dictionary that maps individual sentence to its  aspect\_id(s).
        \item \texttt{aspect\_id2sent}: The inverse mapping of \texttt{sent2aspect\_id}. 
        \item \texttt{aspects}: contains all aspects ids excluding sub-aspects for Query 1. 
        \begin{enumerate}[label={\textbullet}]
            \item \texttt{aspect1\_id}: [\texttt{sub-aspect1\_id}, \texttt{sub-aspect2\_id},\dots]
            \item \texttt{aspect2\_id}:
            \item \dots
        \end{enumerate}
    \end{enumerate}
        \item Query 2:
        \item \dots
    \end{enumerate}
    
    \item \textbf{Corpus}: contains 363,133 Computer Science(CS) paper abstracts.
    \begin{enumerate}[label={\textbullet}]
        \item Abstract 1: 
            \begin{enumerate}[label=-]
        \item \texttt{original\_abstract}: The text of the abstract.
        \item \texttt{masked\_abstract}: The text of the abstract with specialized topic words masked.
        \item \texttt{title}: Title of the paper.
        \item \texttt{url}: Link to the paper.
        \item \texttt{primary\_category}: Primary category of the paper, assigned by arXiv. 
        \item \texttt{categories}: Full list of the categories of the paper, assigned by arXiv. 
        \item \texttt{ss\_id}: The paper's semantic scholar id.
        \item \texttt{incoming\_citations}: List of \texttt{abstract\_id}s citing this paper.
        \item \texttt{outgoing\_citations}: List of \texttt{abstract\_id}s cited by this paper.
        \item \texttt{abstract\_id}: the index 1 is the \texttt{abstract\_id} for this paper.
    \end{enumerate}
    \item Abstract 2:
    \item \dots
    \end{enumerate}

    \item \textbf{Annotation}: contains 83,591 annotated question pairs.
    \begin{enumerate}[label={\textbullet}]
        \item Annotation 1: The first question pair.
            \begin{enumerate}[label=-]
        \item \texttt{aspect\_id}: The id of the \texttt{aspect}/\texttt{sub-aspect} in the question pair.
        \item \texttt{abstract\_id}: The id of the \texttt{abstract} in the question pair.
        \item \texttt{gpt\_response}: The reasoning and decision by ChatGPT for this question pair.
        \item \texttt{score}: The relevance score of the question pair (e.g. 0,1,2).
    \end{enumerate}
    \item Annotation 2:
    \item ...
    \end{enumerate}

    \item \textbf{Test Set}: contains 250 humanly annotated question pairs.
    \begin{enumerate}[label={\textbullet}]
        \item Test Set Question Pair 1:
        \begin{enumerate}[label=-]
            \item \texttt{aspect\_id}: The id of the \texttt{aspect}/\texttt{sub-aspect} in the question pair.
        \item \texttt{abstract\_id}: The id of the \texttt{abstract} in the question pair.
        \item \texttt{human\_annotation}: contains the relevance score assigned by all 3 annotators.
        \begin{enumerate}[label={\textbullet}]
            \item \texttt{annotator\_1}: Score of annotator\_1.
            \item \texttt{annotator\_2}: Score of annotator\_2.
            \item \texttt{annotator\_3}: Score of annotator\_3.
        \end{enumerate}
  
        \end{enumerate}
    \item Test Set Question Pair 2:
    \item \dots 
    \end{enumerate}
\end{enumerate}


\subsection{Statistics of Dataset}
\label{ssec:statistics of dataset}

\begin{table}[ht]
    \begin{minipage}{0.45\linewidth}
        \centering
        \caption{\textcolor{black}{Number of relevant abstracts for each query type. Comp means composite queries. Sub-query stands for two-aspect sub-queries.}}
        \vspace{1em}
        \footnotesize
        \begin{tabular}{@{}cc|cccc@{}}
            \toprule
            &  & Min & Max & Mean & Std \\
            \midrule
            \multirow{6}{*}{Query} & ML & 2 & 84 & 24.35 & 19.47 \\
            & NLP & 0 & 41 & 17.65 & 11.98 \\
            & CV & 0 & 52 & 18.55 & 13.29 \\
            & AI & 1 & 51 & 11.60 & 12.35 \\
            & Comp & 0 & 61 & 9.95 & 14.63 \\
            \midrule
            & All & 0 & 84 & 16.42 & 15.49 \\
            \midrule
            \midrule
            \multirow{6}{*}{Sub-query} & ML & 1 & 105 & 36.40 & 25.34 \\
            & NLP & 0 & 83 & 29.86 & 18.56 \\
            & CV & 1 & 99 & 28.52 & 16.57 \\
            & AI & 0 & 79 & 20.96 & 16.16 \\
            & Comp & 0 & 81 & 23.09 & 21.91 \\
            \midrule
            & All & 0 & 105 & 28.14 & 21.07 \\
            \bottomrule
        \end{tabular}
        \label{tab:statistics_of_relevant_abstracts}
    \end{minipage}%
    \hfill
    \begin{minipage}{0.45\linewidth}
        \centering
        \caption{\textcolor{black}{Normalized score of abstracts for each query type. Compo means composite queries. Sub-query stands for two-aspect sub-queries.}}
        \vspace{1em}
        \footnotesize
        \begin{tabular}{@{}cc|cccc@{}}
            \toprule
            &  & Min & Max & Mean & Std \\
            \midrule
            \multirow{6}{*}{Query} & ML & 0 & 1.778 & 0.665 & 0.377 \\
            & NLP & 0 & 1.786 & 0.609 & 0.361 \\
            & CV & 0 & 1.615 & 0.623 & 0.339 \\
            & AI & 0 & 1.667 & 0.510 & 0.337 \\
            & Comp & 0 & 1.909 & 0.483 & 0.332 \\
            \midrule
            & All & 0 & 1.909 & 0.578 & 0.356 \\
            \midrule
            \midrule
            \multirow{6}{*}{Sub-query} & ML & 0 & 2.000 & 0.679 & 0.478 \\
            & NLP & 0 & 2.000 & 0.638 & 0.453 \\
            & CV & 0 & 2.000 & 0.642 & 0.434 \\
            & AI & 0 & 2.000 & 0.533 & 0.419 \\
            & Comp & 0 & 2.000 & 0.544 & 0.425 \\
            \midrule
            & All & 0 & 2.000 & 0.610 & 0.448 \\
            \bottomrule
        \end{tabular}
        \label{tab:statistics_of_normalized_scores}
    \end{minipage}
\end{table}



\begin{figure}[!htb]
    \centering

\begin{subfigure}[b]{\textwidth}
    \includegraphics[width=0.95\textwidth]{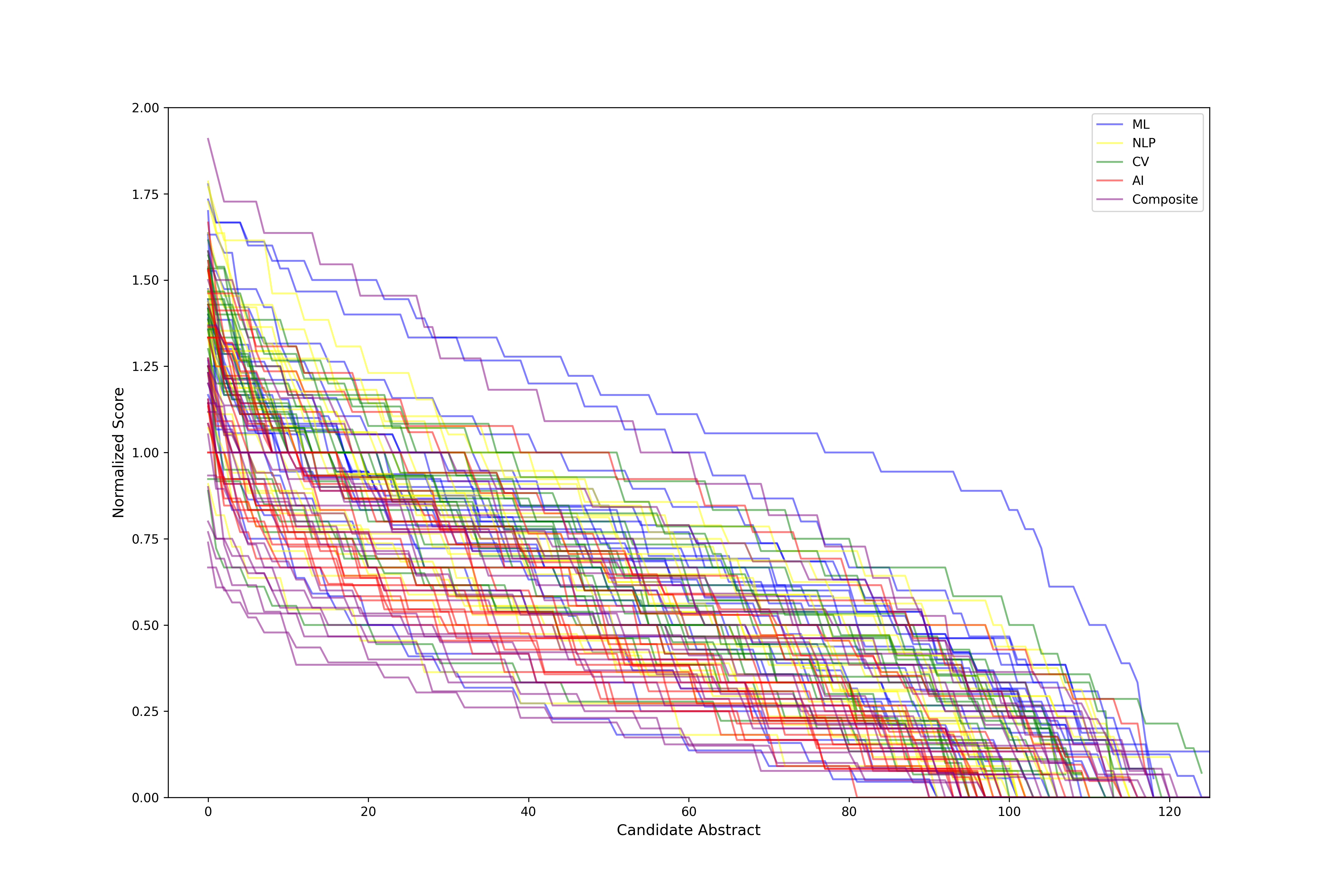}
    \caption{\textcolor{black}{Normalized scores of abstracts in candidate pool per query. }}
\end{subfigure}     

\begin{subfigure}[b]{\textwidth}
    \includegraphics[width=0.95\textwidth]{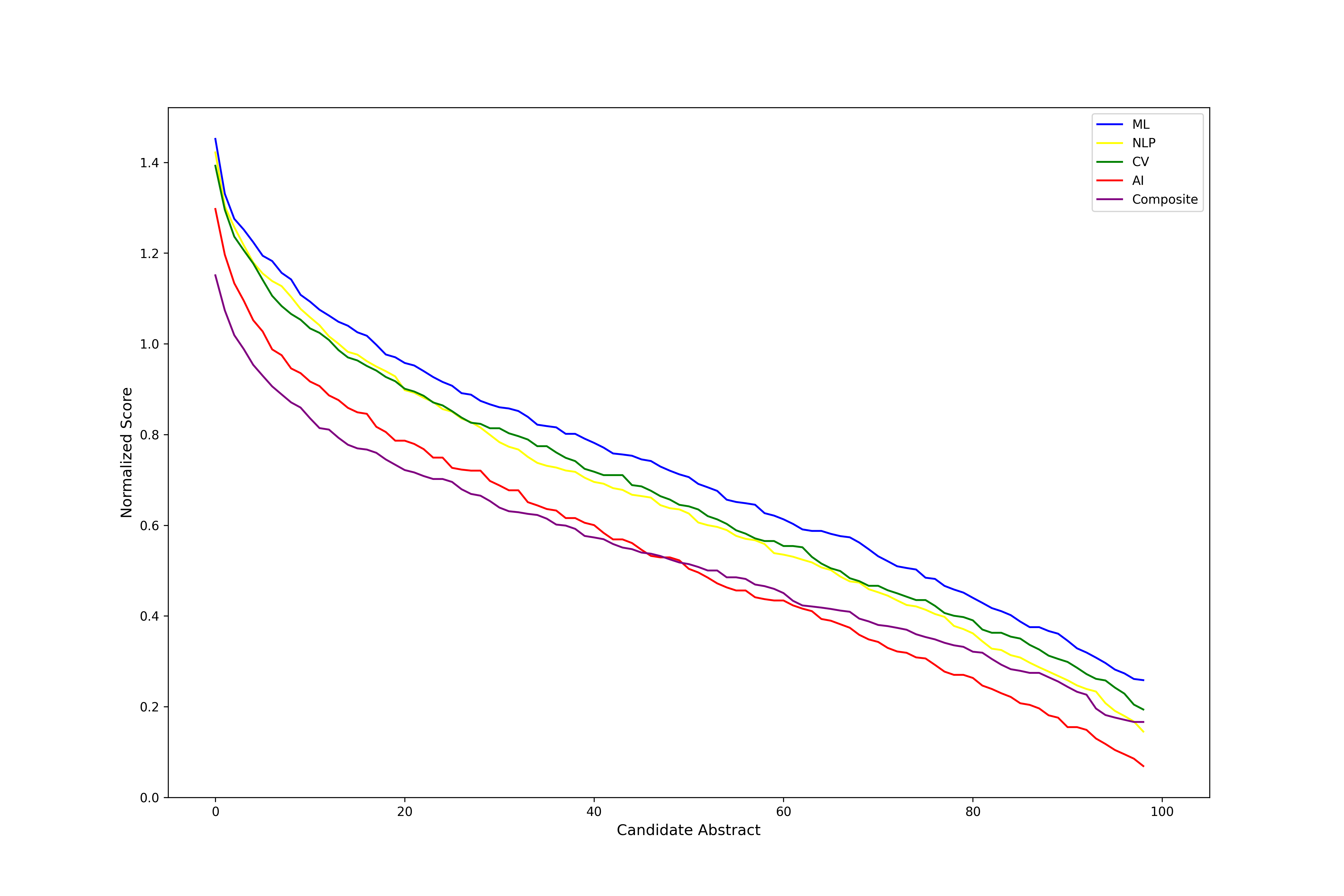}
    \caption{\textcolor{black}{Aggregated normalized scores of abstracts per query type.}}
\end{subfigure}   
        \caption{\textcolor{black}{Each line in (a) represents a query. Each line in (b) represents a query type. The color of the line represents the type of the query. The candidate abstracts are in descending order w.r.t. their normalized scores. }}
        \label{fig:score_of_cand_pool}
    \end{figure}

To better interpret the results of our evaluations and guide future improvements, we provide several analyses of the distribution and relevance of the abstracts in our DORIS-MAE dataset. 

We used Equations \ref{eq:full-q} and \ref{eq:sub-q} to determine the relevance of an abstract to a specific query. We then examined the distribution of the number of relevant abstracts for different queries and considered how this might impact various evaluation metrics.

\textcolor{black}{Table \ref{tab:statistics_of_relevant_abstracts} presents the distribution of relevant abstracts categorized by query type (ML, NLP, CV, AI, and Composite) in our dataset. For each category, we determined the count of relevant abstracts for its respective queries and provided summary statistics of these counts. We find that composite queries have a lower average number of relevant abstracts, implying that such queries might encompass aspects or sub-aspects not extensively covered by papers in their associated pools. This observation, along with the varied number of relevant abstracts per query (ranging from 0 to 84), suggests that the DORIS-MAE dataset encompasses queries of diverse difficulty levels. It's noteworthy that four queries lack any relevant abstracts and were consequently omitted from our evaluation.}

\textcolor{black}{The lower portion of Table \ref{tab:statistics_of_relevant_abstracts} presents the distribution of relevant abstracts for 1003 sub-queries with two aspects. As the table shows, the number of relevant abstracts increases across all five types, suggesting that the sub-query DORIS-MAE task is less challenging than the full-query DORIS-MAE task.}

Next, we examined the distribution of normalized scores (defined as $\frac{S(p_j|Q)}{|Q|}$) for each abstract given a query. We compiled the normalized scores for each abstract in the candidate pools for each query type and report the aggregate statistics in Table \ref{tab:statistics_of_normalized_scores}. Composite queries have the lowest average relevance scores.

Figure \ref{fig:score_of_cand_pool}(a) shows the normalized scores within a candidate pool (ranked in descending order) for a given query. Figure \ref{fig:score_of_cand_pool}(b) aggregates the data from Figure \ref{fig:score_of_cand_pool}(a). As seen in Figure \ref{fig:score_of_cand_pool}(a), there are only a few plateaus (i.e., abstracts with tied scores), which means the normalized scores provide a fine-grained ranking for the candidate pool. This granularity is valuable when evaluating reranking models.

\section{Annotation}
\label{sec:annotation}

\subsection{Annotation Guidelines}
\label{ssec:annotation guidelines}

In Section \ref{sec: anno-gpt framework}, we introduced the grading scale (0-2) for evaluating the relationship between an abstract and an aspect/sub-aspect in a question pair. To facilitate this annotation process, we expand the definition of Level 1 into two sub-categories, creating a four-level grading scale during the annotation phase. Afterward, these levels are mapped back onto the original three-level scale.

The four-level grading scale and their corresponding mappings are as follows:

\begin{enumerate}
    \item \textbf{DISAGREE}: This corresponds to Level 0 in the main paper. The abstract is unrelated or does not provide any help to the key components of the aspect or sub-aspect.
    \item \textbf{DISPUTE}: This is a sub-category of Level 1. The abstract answers some key components and is adaptable to fulfill the aspect/sub-aspect. However, the abstract does not directly address the aspect/sub-aspect.
    \item \textbf{AGREE}: This is another sub-category of Level 1. The abstract directly addresses only a portion of the aspect/sub-aspect, but not in its entirety.
    \item \textbf{CONCUR}: This corresponds to Level 2. The abstract directly answers all the key components of the aspect or sub-aspect.
\end{enumerate}

These labels "DISAGREE", "DISPUTE", "AGREE", and "CONCUR" serve to add semantic meaning to the grading levels, aiding annotators during the annotation process.

For consistency, "DISPUTE" and "AGREE" are treated as equal in weight, representing different scenarios under Level 1, but neither is considered stronger or weaker than the other.

During the annotation process, annotators must follow a set of guidelines designed to maintain consistency and minimize bias. These guidelines cover aspects like time management, communication among annotators (which is explicitly forbidden), use of additional resources, the process of question analysis, and the identification of key components. Details of these guidelines can be found in the following section.

    

\begin{enumerate}
    \item \textbf{Resource usage}: Given the technical nature of the abstracts, annotators may occasionally need to look up unfamiliar terms. They can use traditional search engines for these instances. However, they are strictly prohibited from using any large language model (LLM) interfaces or search engines powered by LLMs.

    \item \textbf{Annotation process}: Annotators are required to read the entire abstract and question at least once before making a decision. They need to refresh their understanding of the four grading levels in their minds before finalizing their decision. On average, this process takes about 4 minutes, but no strict time limit is enforced. During the annotation process, annotators are strictly and explicitly prohibited from communicating in any way. 

    \item \textbf{Key component identification}: The grading scale revolves around the concept of "key components". It's up to the annotators, leveraging their research experience, to identify what constitutes as a key component in an aspect/sub-aspect. If there are multiple key components (e.g., 2 or 3), they are expected to monitor all of these components and verify whether the abstract mentions them.

    \item \textbf{Inference rules}: Annotators can leverage their expert knowledge to draw reasonable inferences from the abstract's content with respect to the key components. However, they should avoid unnecessary complications in the correlation between the aspect and the abstract. Each inference made must have a clear, straightforward justification. The aim here is to ensure accuracy without over-complicating the annotation process.

\end{enumerate}

\subsection{Examples for Annotations}
\label{ssec:examples for annotations}

In this section, we provide examples of aspect/abstract pairs with different relevance scores.

\subsubsection{Examples with relevance score 0}

\begin{itemize}[label={}]
    \item 

\begin{quote}
    \textit{Aspect/Sub-aspect}: The paper mentions the possibility that the labeling process introduces unwanted randomness or noise.\\

    \textit{Abstract}: Recent approaches in literature have exploited the multi-modal information in documents (text, layout, image) to serve specific downstream document tasks. However, they are limited by their - (i) inability to learn cross-modal representations across text, layout and image dimensions for documents and (ii) inability to process multi-page documents. Pre-training techniques have been shown in Natural Language Processing (NLP) domain to learn generic textual representations from large unlabelled datasets, applicable to various downstream NLP tasks. In this paper, we propose a multi-task learning-based framework that utilizes a combination of self-supervised and supervised pre-training tasks to learn a generic document representation applicable to various downstream document tasks. Specifically, we introduce Document Topic Modelling and Document Shuffle Prediction as novel pre-training tasks to learn rich image representations along with the text and layout representations for documents. We utilize the Longformer network architecture as the backbone to encode the multi-modal information from multi-page documents in an end-to-end fashion. We showcase the applicability of our pre-training framework on a variety of different real-world document tasks such as document classification, document information extraction, and document retrieval. We evaluate our framework on different standard document datasets and conduct exhaustive experiments to compare performance against various ablations of our framework and state-of-the-art baselines. 
\end{quote}
\begin{quote}
    \textit{Human reasoning}: The above abstract is not related to the aspect.
\end{quote}
 \hspace{0.5em} \dotfill
\item 

\begin{quote}
    \textit{Aspect/Sub-aspect}: The document should use semi-definite programming to estimate error bound.\\

    \textit{Abstract}: We develop methodology for estimation and inference using machine learning to enrich economic models. Our framework takes a standard economic model and recasts the parameters as fully flexible nonparametric functions, to capture the rich heterogeneity based on potentially high dimensional or complex observable characteristics. These "parameter functions" retain the interpretability, economic meaning, and discipline of classical parameters. Deep learning is particularly well-suited to structured modeling of heterogeneity in economics. We show how to design the network architecture to match the structure of the economic model, delivering novel methodology that moves deep learning beyond prediction. We prove convergence rates for the estimated parameter functions. These functions are the key inputs into the finite-dimensional parameter of inferential interest. We obtain inference based on a novel influence function calculation that covers any second-stage parameter and any machine-learning-enriched model that uses a smooth per-observation loss function. No additional derivations are required. The score can be taken directly to data, using automatic differentiation if needed. The researcher need only define the original model and define the parameter of interest. A key insight is that we need not write down the influence function in order to evaluate it on the data. Our framework gives new results for a host of contexts, covering such diverse examples as price elasticities, willingness-to-pay, and surplus measures in binary or multinomial choice models, effects of continuous treatment variables, fractional outcome models, count data, heterogeneous production functions, and more. We apply our methodology to a large scale advertising experiment for short-term loans. We show how economically meaningful estimates and inferences can be made that would be unavailable without our results.
\end{quote}
\begin{quote}
    \textit{Human reasoning}: The above abstract is not related to the aspect.
\end{quote}

\end{itemize}

\subsubsection{Examples with relevance score 1}
\begin{itemize}[label={}]
    \item 

\begin{quote}
    \textit{Aspect/Sub-aspect}: The paper should introduce a system that can represent the possible options and consequences in decision making using actions and states.\\

    \textit{Abstract}: Decision-making is often dependent on uncertain data, e.g. data associated with confidence scores or probabilities. We present a comparison of different information presentations for uncertain data and, for the first time, measure their effects on human decision-making. We show that the use of Natural Language Generation (NLG) improves decision-making under uncertainty, compared to state-of-the-art graphical-based representation methods. In a task-based study with 442 adults, we found that presentations using NLG lead to 24\% better decision-making on average than the graphical presentations, and to 44\% better decision-making when NLG is combined with graphics. We also show that women achieve significantly better results when presented with NLG output (an 87\% increase on average compared to graphical presentations).
\end{quote}
\begin{quote}
   \textit{Human reasoning}: The above abstract explicitly discuss decision-making. However, the abstract does not mention anything about actions and states. Thus, some key components are missing. More specifically, it belongs to the class "Agree" under Level 1.  
\end{quote}
 \hspace{0.5em} \dotfill
\item 
\begin{quote}
\textit{Aspect/Sub-aspect}: The document should introduce a system that can retrieve fine-grained video of actions from longer videos and match these actions to the text description or instruction.\\

\textit{Abstract}:
In this paper, we propose a new approach for retrieval of video segments using natural language queries. Unlike most previous approaches such as concept-based methods or rule-based structured models, the proposed method uses image captioning model to construct sentential queries for visual information. In detail, our approach exploits multiple captions generated by visual features in each image with `Densecap'. Then, the similarities between captions of adjacent images are calculated, which is used to track semantically similar captions over multiple frames. Besides introducing this novel idea of 'tracking by captioning', the proposed method is one of the first approaches that uses a language generation model learned by neural networks to construct semantic query describing the relations and properties of visual information. To evaluate the effectiveness of our approach, we have created a new evaluation dataset, which contains about 348 segments of scenes in 20 movie-trailers. Through quantitative and qualitative evaluation, we show that our method is effective for retrieval of video segments using natural language queries.
\end{quote}
\begin{quote}
    \textit{Human reasoning}: The above abstract mentions retrieval of video segments using natural language queries, which is related to the aspect. However, another key component of the abstract is "action". Although it is very natural to infer that the system in the document can retrieve video of actions, it is not explicitly mentioned in the abstract. More specifically, it belongs to the class "Dispute" under Level 1. 
\end{quote}
\end{itemize}

\subsubsection{Examples with relevance score 2}
\begin{itemize}[label={}]
    \item 

\begin{quote}
\textit{Aspect/Sub-aspect}: The paper should mention chemistry or chemistry related knowledge.\\

\textit{Abstract}: Modern astronomical surveys are observing spectral data for millions of stars. These spectra contain chemical information that can be used to trace the Galaxy's formation and chemical enrichment history. However, extracting the information from spectra, and making precise and accurate chemical abundance measurements are challenging. Here, we present a data-driven method for isolating the chemical factors of variation in stellar spectra from those of other parameters (i.e. \textbackslash teff, \textbackslash logg, eh). This enables us to build a spectral projection for each star with these parameters removed. We do this with no ab initio knowledge of elemental abundances themselves, and hence bypass the uncertainties and systematics associated with modeling that rely on synthetic stellar spectra. To remove known non-chemical factors of variation, we develop and implement a neural network architecture that learns a disentangled spectral representation. We simulate our recovery of chemically identical stars using the disentangled spectra in a synthetic APOGEE-like dataset. We show that this recovery declines as a function of the signal to noise ratio, but that our neural network architecture outperforms simpler modeling choices. Our work demonstrates the feasibility of data-driven abundance-free chemical tagging.
\end{quote}
\begin{quote}
    \textit{Human reasoning}: The above abstract explicitly mention "recovery of chemically identical stars" which is related to chemistry.
\end{quote}
\hspace{0.5em} \dotfill
\item 
\begin{quote}
\textit{Aspect/Sub-aspect}: The paper should provide robust safeguards to sensitive patient data.\\

\textit{Abstract}: In survival analysis, regression models are used to understand the effects of explanatory variables (e.g., age, sex, weight, etc.) to the survival probability. However, for sensitive survival data such as medical data, there are serious concerns about the privacy of individuals in the data set when medical data is used to fit the regression models. The closest work addressing such privacy concerns is the work on Cox regression which linearly projects the original data to a lower dimensional space. However, the weakness of this approach is that there is no formal privacy guarantee for such projection. In this work, we aim to propose solutions for the regression problem in survival analysis with the protection of differential privacy which is a golden standard of privacy protection in data privacy research. To this end, we extend the Output Perturbation and Objective Perturbation approaches which are originally proposed to protect differential privacy for the Empirical Risk Minimization (ERM) problems. In addition, we also propose a novel sampling approach based on the Markov Chain Monte Carlo (MCMC) method to practically guarantee differential privacy with better accuracy. We show that our proposed approaches achieve good accuracy as compared to the non-private results while guaranteeing differential privacy for individuals in the private data set.
\end{quote}
\begin{quote}
    \textit{Human reasoning}: The above abstract explicitly discuss differential privacy" and sensitive private medical data, which satisfies all components in the abstract.
\end{quote}
\end{itemize}

\subsection{Prompt Engineering}
\label{ssec:prompt engineering}

The prompt strategy was optimized on the development set, and was designed with an emphasis on clarity and conciseness. It aimed to elicit detailed reasoning and a definitive final decision from ChatGPT.

Key to the final prompt design was the selection of labels for the grading options: "Disagree," "Dispute," "Agree," and "Concur." These labels were chosen to reflect the different grading scale levels defined in Section \ref{ssec:annotation guidelines}.


These labels, along with their detailed definitions, are included in the prompt. They serve to guide ChatGPT in its decision-making process by providing consistent terminology for each grading level.
The final prompt used in the hypothesis testing stage is displayed below.

\begin{quote}

    \texttt{You are a reference librarian. A user wants to find a scientific paper that satisfies a particular requirement. Based on the abstract of the paper, you will determine whether the requirement is met.} 
    \vspace{0.2cm}
    
    \texttt{Here is the user requirement: \{aspect\}}
    
    \texttt{Here is the abstract: \{abstract\}}

    \vspace{0.2cm}
    \texttt{Explain whether the abstract satisfies the user requirement. Think step by step, showing all of your reasoning. Use the following *rules* to determine whether the requirement is satisfied:} 
    \vspace{0.2cm}
    
    \texttt{Disagree means the abstract does not address the user requirement;} 
    
    \texttt{Dispute means the abstract does not directly address the user requirement, but is adaptable to fulfill the user’s need;} 
    
    \texttt{Concur means the abstract directly and completely addresses the user requirement;} 
    
    \texttt{Agree means the abstract directly addresses only a portion of the user requirement, but not in its entirety.}

    \vspace{0.2cm}
    \texttt{Think carefully about the aforementioned *rules* to decide whether you [DISAGREE, DISPUTE, AGREE, CONCUR] based on the definition of these 4 options. After you have provided your reasoning, state your final selection in *all capital* letters at the *end* of your response.}
\end{quote}

This prompt produced the best results on the development set, with the distribution of its responses closely matching that of human annotators. Based on these results, we proceeded to use this prompt in the testing process.

\subsection{Error Analysis for ChatGPT's Annotation}
\label{ssec:error analysis for chatgpt's annotation}

Though the final prompt yielded results comparable to human annotators, we aimed to identify potential inconsistencies in ChatGPT's responses that slightly deviate from humans' answers. We have organized them into several categories:

\subsubsection{Definition Inconsistency}

Test Case Prompt: 
\begin{quote}
    You are a reference librarian. A user wants to find a scientific paper that satisfies a particular requirement. Based on the abstract of the paper, you will determine whether the requirement is met. 

Here is the user requirement: The model in the document should have the ability to discover the geometry and global structure of the face

Here is the abstract: Facial image super-resolution (SR) is an important preprocessing for facial
image analysis, face recognition, and image-based 3D face reconstruction.
Recent convolutional neural network (CNN) based method has shown excellent
performance by learning mapping relation using pairs of low-resolution (LR) and
high-resolution (HR) facial images. However, since the HR facial image
reconstruction using CNN is conventionally aimed to increase the PSNR and SSIM
metrics, the reconstructed HR image might not be realistic even with high
scores. An adversarial framework is proposed in this study to reconstruct the
HR facial image by simultaneously generating an HR image with and without blur.
First, the spatial resolution of the LR facial image is increased by eight
times using a five-layer CNN. Then, the encoder extracts the features of the
up-scaled image. These features are finally sent to two branches (decoders) to
generate an HR facial image with and without blur. In addition, local and
global discriminators are combined to focus on the reconstruction of HR facial
structures. Experiment results show that the proposed algorithm generates a
realistic HR facial image. Furthermore, the proposed method can generate a
variety of different facial images.

Explain whether the abstract satisfies the user requirement. Think step by step, showing all of your reasoning. Use the following *rules* to determine whether the requirement is satisfied: 
disagree means the abstract does not address the user requirement; 
dispute means the abstract does not directly address the user requirement, but is adaptable to fulfill the user’s need; 
concur means the abstract directly and completely addresses the user requirement; 
agree means the abstract directly addresses only a portion of the user requirement, but not in its entirety. Think carefully about the aforementioned *rules* to decide whether you [DISAGREE, DISPUTE, AGREE, CONCUR] based on the definition of these 4 options. After you have provided your reasoning, state your final selection in *all capital* letters at the *end* of your response.
\end{quote}
ChatGPT's Reasoning:
\begin{quote}
    The user requirement is that the model in the document should have the ability to discover the geometry and global structure of the face. The abstract discusses a proposed algorithm for facial image super-resolution using an adversarial framework that generates a realistic HR facial image. The algorithm uses local and global discriminators to focus on the reconstruction of HR facial structures. 
Based on this information, I CONCUR that the abstract satisfies the user requirement. The proposed algorithm is designed to reconstruct the HR facial image by simultaneously generating an HR image with and without blur. The local and global discriminators are combined to focus on the reconstruction of HR facial structures, which implies that the model has the ability to discover the geometry and global structure of the face. Therefore, the abstract directly and completely addresses the user requirement.
\end{quote}

\paragraph{Error Analysis:}
Despite the abstract's implied ability to discern facial structures, it doesn't state this explicitly. However, even after recognizing the ability as "implied," ChatGPT marked the abstract as directly satisfying the requirement. This indicates occasional misalignment between ChatGPT's reasoning and the labeling criteria.

\subsubsection{Inference Inconsistency}
Test Case Prompt:

\begin{quote}
    You are a reference librarian. A user wants to find a scientific paper that satisfies a particular requirement. Based on the abstract of the paper, you will determine whether the requirement is met. 

Here is the user requirement: The document should introduce a model that can take human faces with stripe patterns and accurately remove the stripe pattern.

Here is the abstract: In this paper, we propose a robust 3D face recognition system which can
handle pose as well as occlusions in real world. The system at first takes as
input, a 3D range image, simultaneously registers it using ICP(Iterative
Closest Point) algorithm. ICP used in this work, registers facial surfaces to a
common model by minimizing distances between a probe model and a gallery model.
However the performance of ICP relies heavily on the initial conditions. Hence,
it is necessary to provide an initial registration, which will be improved
iteratively and finally converge to the best alignment possible. Once the faces
are registered, the occlusions are automatically extracted by thresholding the
depth map values of the 3D image. After the occluded regions are detected,
restoration is done by Principal Component Analysis (PCA). The restored images,
after the removal of occlusions, are then fed to the recognition system for
classification purpose. Features are extracted from the reconstructed
non-occluded face images in the form of face normals. The experimental results
which were obtained on the occluded facial images from the Bosphorus 3D face
database, illustrate that our occlusion compensation scheme has attained a
recognition accuracy of 91.30

Explain whether the abstract satisfies the user requirement. Think step by step, showing all of your reasoning. Use the following *rules* to determine whether the requirement is satisfied: 
disagree means the abstract does not address the user requirement; 
dispute means the abstract does not directly address the user requirement, but is adaptable to fulfill the user’s need; 
concur means the abstract directly and completely addresses the user requirement; 
agree means the abstract directly addresses only a portion of the user requirement, but not in its entirety. Think carefully about the aforementioned *rules* to decide whether you [DISAGREE, DISPUTE, AGREE, CONCUR] based on the definition of these 4 options. After you have provided your reasoning, state your final selection in *all capital* letters at the *end* of your response.
\end{quote}

ChatGPT's Reasoning:
\begin{quote}
    The abstract DISAGREES with the user requirement. The paper proposes a 3D face recognition system that can handle pose and occlusions in real-world scenarios. The system takes a 3D range image as input, registers it using the ICP algorithm, and extracts occlusions by thresholding the depth map values of the 3D image. However, the paper does not introduce a model that can remove stripe patterns from human faces. Therefore, the abstract does not address the user requirement.
\end{quote}
\paragraph{Error Analysis:}
The human annotators labeled this question as "Dispute" due to the mention of occlusion detection, removal, and restoration in the abstract, inferring that "stripe patterns" could fall under the broader category of "occlusions". However, ChatGPT did not make the same inference, indicating a different style of reasoning.

\subsubsection{Key Component Inconsistency}
Test Case Prompt:
\begin{quote}
    You are a reference librarian. A user wants to find a scientific paper that satisfies a particular requirement. Based on the abstract of the paper, you will determine whether the requirement is met. 

Here is the user requirement: The navigation agent model in the document should have the ability to be trained on data with varying levels of augmentation

Here is the abstract: Localization is the problem of estimating the location of an autonomous agent
from an observation and a map of the environment. Traditional methods of
localization, which filter the belief based on the observations, are
sub-optimal in the number of steps required, as they do not decide the actions
taken by the agent. We propose "Active Neural Localizer", a fully
differentiable neural network that learns to localize accurately and
efficiently. The proposed model incorporates ideas of traditional
filtering-based localization methods, by using a structured belief of the state
with multiplicative interactions to propagate belief, and combines it with a
policy model to localize accurately while minimizing the number of steps
required for localization. Active Neural Localizer is trained end-to-end with
reinforcement learning. We use a variety of simulation environments for our
experiments which include random 2D mazes, random mazes in the Doom game engine
and a photo-realistic environment in the Unreal game engine. The results on the
2D environments show the effectiveness of the learned policy in an idealistic
setting while results on the 3D environments demonstrate the model's capability
of learning the policy and perceptual model jointly from raw-pixel based RGB
observations. We also show that a model trained on random textures in the Doom
environment generalizes well to a photo-realistic office space environment in
the Unreal engine.

Explain whether the abstract satisfies the user requirement. Think step by step, showing all of your reasoning. Use the following *rules* to determine whether the requirement is satisfied: 
disagree means the abstract does not address the user requirement; 
dispute means the abstract does not directly address the user requirement, but is adaptable to fulfill the user’s need; 
concur means the abstract directly and completely addresses the user requirement; 
agree means the abstract directly addresses only a portion of the user requirement, but not in its entirety. Think carefully about the aforementioned *rules* to decide whether you [DISAGREE, DISPUTE, AGREE, CONCUR] based on the definition of these 4 options. After you have provided your reasoning, state your final selection in *all capital* letters at the *end* of your response.
\end{quote}

ChatGPT's Reasoning:
\begin{quote}
    The abstract does not directly address the user requirement. The abstract discusses a neural network model for localization of an autonomous agent in various environments. The model is trained with reinforcement learning and is able to learn the policy and perceptual model jointly from raw-pixel based RGB observations. However, there is no mention of the navigation agent model being trained on data with varying levels of augmentation. Therefore, I DISAGREE that the abstract satisfies the user requirement.
\end{quote}
\paragraph{Error Analysis:}

Human annotator 1 assigned the label "Dispute" and human annotator 2 assigned the label "Disagree", while ChatGPT assigned the label "Disagree". The difference between ChatGPT and human annotator 1 in this case comes from their different definition of key components in the aspect. Human annotator 1 believes the navigation agent model itself is a key component, thus by mentioning the localization problem of an autonomous agent, the abstract already indirectly implies the navigation agent model. However, both ChatGPT and annotator 2 believes navigation agent that can be trained on data with varying levels of augmentation is the only key component in the aspect. Therefore, since data with varying level of augmentation is not mentioned in the abstract, both of them assigned "Disagree" indicating the abstract is unrelated to the only key component in the aspect. This form of inconsistency is the most common form of inconsistencies we found in ChatGPT's reasoning versus some human's reasoning. However, since humans also disagree on what constitutes as key components occasionally, this form of key component inconsistency cannot be completely avoided even when ChatGPT exhibits comparable performance as humans.  

\subsection{Annotation Scalability}
\label{ssec:annotation scalability}

The Anno-GPT framework, as detailed in Section \ref{sec: anno-gpt framework}, employs ChatGPT for automated annotations, contingent on its performance being comparable to human experts during the hypothesis testing stage. This system scales linearly with the number of complex queries, aspects/sub-aspects, and the size of candidate pools per query, enabling larger datasets for training and testing purposes. For instance, with ten API keys, a multi-threaded program could create a training-sized dataset of 10,000 complex queries, each with a candidate pool of 100 abstracts, in approximately 240 hours.

Creating a DORIS-MAE dataset for another field, such as quantum computing or biomedical engineering, involves several preparatory stages. Initial complex queries (e.g., 50 queries) need to be either expertly crafted or gathered from online technical discussions in the field.
Next, the complex queries must be broken down into a list of aspects and sub-aspects. This process could be carried out by an expert in the field or a proficient annotator.

Finally, it is critical to fine-tune the final prompt and verify ChatGPT's competence in the new field of research. As per our annotation guidelines, this stage involves three expert annotators, each dedicating an average of 20 hours. This step constitutes the largest fixed cost of applying Anno-GPT to a new field, but it's a necessary investment for ensuring dataset quality. However, once a language model such as ChatGPT is confirmed to perform at a comparable level to human experts, the Anno-GPT framework can then be used to create larger datasets in the domain.

\section{Retrieval Results}
\label{sec:experiment results}
This section presents the full results of benchmarking sixteen models (ada-002 \cite{openaiada}, E5-Large-v2 \cite{e5largev2}, LLAMA \cite{llama}, Text-Supervised (TS)-ASPIRE \cite{aspire}, Optimal-Transport (OT)-ASPIRE \cite{aspire}, SPECTER \cite{specter}, Sentence-BERT \cite{sentbert}, RocketQA-v2 \cite{rocketqa_v2}, ANCE FirstP \cite{ance}, SimLM \cite{simlm}, SPLADE-v2 \cite{spladev2}, ColBERT-v2 \cite{colbertv2}, SciBERT \cite{scibert}, ERNIE-Search \cite{ernie}, BM25 \cite{bm25}, TF-IDF \cite{tfidf}) discussed in Section \ref{sec: related work} on the DORIS-MAE dataset. For SciBERT and SPECTER, we trained them on our 360k CS corpus with citation signals (i.e. in domain) and denoted their best trained version as SciBERT\textsubscript{ID} and SPECTER\textsubscript{ID}.

\subsection{Metrics}
\label{ssec:metric introduction}
In our experiments, we used several metrics which are commonly used in the IR/NIR literature for evaluating ranking and retrieval models. These include Recall@5 (R@5), Recall@20 (R@20), R-Precision (RP), NDCG\textsubscript{10\%}, $\textrm{NDCG}^{\textrm{exp}}_{10\%}$ \cite{ndcg_exp}, Mean Reciprocal Rank@10 (MRR@10), and Mean Average Precision (MAP). An abstract was categorized as relevant when its score from Equation \ref{eq:full-q} $\frac{S(p_j|Q)}{|Q|}\geq 1$. Note that NDCG\textsubscript{10\%} is excluded from the main paper due to an atypically high random baseline value of \textcolor{black}{48.71\%}.

\subsection{Query Level Benchmark Results}
\label{ssec:query level benchmark results}
\subsubsection{Whole Query Embedding}
\label{ssec:read the whole query once}

In Table \ref{tab: Query-level read all appendix}, we showcase the results obtained when models (ada-002, E5-Large-v2, LLAMA, SPECTER, SentBERT, RocketQA-v2, ANCE, SPLADE-v2, ColBERT-v2, SciBERT, ERNIE, BM25, TF-IDF) embed the entire query as a single vector and embed the entire abstract as a single vector. The relevance between the query and the abstract is determined by calculating the cosine similarity or L2 distance (with SPECTER using L2 distance as recommended by \citet{specter}) between the query vector and the abstract vector.

In Table \ref{tab: Query-level read all appendix}, we see the general-purposed text embedding models (ada-002, E5-Large-v2, LLAMA, SPECTER) (that are not restricted by the size of context window) typically outperform specialized NIR models (RocketQA-v2, SimLM, SPLADE-v2, ColBERT-v2). 
In-domain fine-tuning improves performance for SciBERT but shows mixed results for SPECTER.

\begin{table}[ht]
\centering
\footnotesize
\caption{Models' query-level ranking performance (embedding query and abstract as single vectors)}
\vspace{1em}
\setlength{\tabcolsep}{3pt}
\begin{tabular}{lcccccccc}
\toprule
Method & R@5 & R@20 & RP & $\textrm{NDCG}\textsubscript{10\%}$ & $\textrm{NDCG}^{\textrm{exp}}_{10\%}$ & MRR@10 & MAP \\
\midrule
random & 4.41 & 18.48 & 16.29 & 48.71 & 7.31 & 3.59 & 19.63 \\
\midrule
ada-002 & 15.38 \scriptsize{$\pm$ 1.95} & 42.84 \scriptsize{$\pm$ 2.53} & 35.81 
\scriptsize{$\pm$ 2.67} & \textbf{72.48}
\scriptsize{$\pm$ 1.10} & \textbf{27.46} \scriptsize{$\pm$ 2.48} & \textbf{19.88} \scriptsize{$\pm$ 3.21} & 40.37 \scriptsize{$\pm$ 2.55}\\
E5-L-v2 & \textbf{16.51} \scriptsize{$\pm$ 2.05} & \textbf{43.77} \scriptsize{$\pm$ 2.14} & \textbf{37.46}
\scriptsize{$\pm$ 2.44} & 70.11
\scriptsize{$\pm$ 0.97} & 25.90 \scriptsize{$\pm$ 2.15} & 14.85 \scriptsize{$\pm$ 2.73} & \textbf{40.49} \scriptsize{$\pm$ 2.32} \\
LLAMA & 12.74 \scriptsize{$\pm$ 1.82} & 34.51 \scriptsize{$\pm$ 2.36} & 28.33 \scriptsize{$\pm$ 2.14} & 64.75 \scriptsize{$\pm$ 1.04} & 16.65 \scriptsize{$\pm$ 1.68} & 11.78 \scriptsize{$\pm$ 2.45} & 31.29 \scriptsize{$\pm$ 1.99} \\
SPLADE-v2\cite{simcse} & 14.78 \scriptsize{$\pm$ 1.89} & 40.14 \scriptsize{$\pm$ 2.33} &31.65 \scriptsize{$\pm$ 2.38} & 71.29 \scriptsize{$\pm$ 0.89}   & 26.08 \scriptsize{$\pm$ 2.00} & 17.82 \scriptsize{$\pm$  2.99} & 37.23 \scriptsize{$\pm$ 2.26}\\
SPECTER-v2 & 14.50 \scriptsize{$\pm$ 2.15} & 43.36 \scriptsize{$\pm$ 2.50} & 33.41 \scriptsize{$\pm$ 2.33} & 70.19 \scriptsize{$\pm$ 0.98} & 25.65 \scriptsize{$\pm$ 2.23} & 17.19 \scriptsize{$\pm$ 2.96} & 37.12 \scriptsize{$\pm$ 2.10} \\
SPECTER & 13.34 \scriptsize{$\pm$ 1.92} & 40.25 \scriptsize{$\pm$ 2.60} & 32.66 \scriptsize{$\pm$ 2.55} & 68.06 \scriptsize{$\pm$ 1.15} & 24.43 \scriptsize{$\pm$ 2.35} & 18.23 \scriptsize{$\pm$ 2.99} & 35.75 \scriptsize{$\pm$ 2.46} \\
SPECTER\textsubscript{ID} & 13.32 \scriptsize{$\pm$ 1.76} & 42.52 \scriptsize{$\pm$ 2.37} & 31.55 \scriptsize{$\pm$ 2.28} & 69.34
\scriptsize{$\pm$ 1.09} & 21.27 \scriptsize{$\pm$ 2.03} & 14.48 \scriptsize{$\pm$ 2.78} & 36.02 \scriptsize{$\pm$ 2.19}\\
SentBERT & 14.39 \scriptsize{$\pm$ 1.97} & 42.42 \scriptsize{$\pm$ 2.45} & 32.63 \scriptsize{$\pm$ 2.60} & 69.70 \scriptsize{$\pm$ 1.11} & 22.94 \scriptsize{$\pm$ 2.21} & 13.28 \scriptsize{$\pm$ 2.71} & 37.47 \scriptsize{$\pm$ 2.34}\\
RocketQA-v2 & 10.59 \scriptsize{$\pm$ 1.20} & 33.72 \scriptsize{$\pm$ 2.21} & 28.42 \scriptsize{$\pm$ 2.18} & 66.17 \scriptsize{$\pm$ 1.17}& 18.57 \scriptsize{$\pm$ 2.01} & 11.51 \scriptsize{$\pm$ 2.49} & 32.57 \scriptsize{$\pm$ 2.14} \\
ANCE\textsubscript{FirstP} & 11.10 \scriptsize{$\pm$ 1.53} & 32.09 \scriptsize{$\pm$ 2.09} & 28.60 \scriptsize{$\pm$ 2.14} & 64.81 \scriptsize{$\pm$ 0.99} & 17.18 \scriptsize{$\pm$ 1.74}   & 9.07 \scriptsize{$\pm$ 1.93} & 30.72 \scriptsize{$\pm$ 2.02} \\
SimLM & 12.68 \scriptsize{$\pm$ 1.77} & 35.67 \scriptsize{$\pm$ 2.49} & 28.90 \scriptsize{$\pm$ 2.42} & 66.31 \scriptsize{$\pm$ 1.10} & 18.91 \scriptsize{$\pm$ 1.86} & 11.29 \scriptsize{$\pm$ 2.44} & 33.06 \scriptsize{$\pm$ 2.34} \\
SimCSE & 14.90 \scriptsize{$\pm$ 1.89} & 42.62 \scriptsize{$\pm$ 2.40} & 35.27 \scriptsize{$\pm$ 2.34} & 70.18 \scriptsize{$\pm$ 1.01} & 26.88 \scriptsize{$\pm$ 2.36} & 21.19 \scriptsize{$\pm$ 3.47} & 39.02 \scriptsize{$\pm$ 2.35}\\

ColBERT-v2 & 8.45 \scriptsize{$\pm$ 1.46} & 27.86 \scriptsize{$\pm$ 2.29} & 22.33 \scriptsize{$\pm$ 2.01} & 59.81 \scriptsize{$\pm$ 1.16} & 12.57 \scriptsize{$\pm$ 1.71} & 6.69 \scriptsize{$\pm$ 2.15} & 25.80 \scriptsize{$\pm$ 1.83} \\
SciBERT & 5.13 \scriptsize{$\pm$ 1.25} & 17.99 \scriptsize{$\pm$ 1.69} & 17.13 \scriptsize{$\pm$ 1.88} & 47.28 \scriptsize{$\pm$ 1.17} & 7.50 \scriptsize{$\pm$ 1.34} & 3.41 \scriptsize{$\pm$ 1.57} & 20.34 \scriptsize{$\pm$ 1.64} \\
SciBERT\textsubscript{ID} & 12.86 \scriptsize{$\pm$ 1.86} & 35.53 \scriptsize{$\pm$ 2.23} & 27.54 \scriptsize{$\pm$ 2.29} & 64.55 \scriptsize{$\pm$ 1.17} & 17.24 \scriptsize{$\pm$ 1.64} & 8.57 \scriptsize{$\pm$ 1.79} & 31.50 \scriptsize{$\pm$ 2.02} \\
ERNIE & 6.49 \scriptsize{$\pm$ 0.94} & 22.58 \scriptsize{$\pm$ 1.72} & 20.18 \scriptsize{$\pm$ 1.82} & 54.91 \scriptsize{$\pm$ 1.17} & 9.66 \scriptsize{$\pm$ 1.18} & 3.77 \scriptsize{$\pm$ 1.06} & 22.71 \scriptsize{$\pm$ 1.65} \\
BM25 & 8.47 \scriptsize{$\pm$ 1.80} & 30.50 \scriptsize{$\pm$ 2.38} & 21.94 \scriptsize{$\pm$ 2.03} & 51.46 \scriptsize{$\pm$ 1.40} & 13.23 \scriptsize{$\pm$ 1.97} & 9.19 \scriptsize{$\pm$ 2.46} & 25.99 \scriptsize{$\pm$ 1.68} \\
TF-IDF & 10.71 \scriptsize{$\pm$ 1.48} & 29.22 \scriptsize{$\pm$ 2.25} & 24.79 \scriptsize{$\pm$ 2.06} & 58.31 \scriptsize{$\pm$ 1.12} & 18.25 \scriptsize{$\pm$ 2.01} & 12.41 \scriptsize{$\pm$ 2.53} & 28.77 \scriptsize{$\pm$ 1.81}\\
\bottomrule
\end{tabular}
\label{tab: Query-level read all appendix}
\end{table}

\subsubsection{Sentence-by-Sentence Embedding}
\label{ssec: read sentence by sentence}

For fairness, we adopted an alternative approach for models like RocketQA-v2 \cite{rocketqa_v2} and ColBERT-v2 \cite{colbertv2} that are trained on shorter queries such as in MS MARCO \cite{msmarco} and NQ \cite{NQ}. In Table \ref{tab:ranking_query_level_sent_by_sent_appendix}, we show the results when these models process each sentence in the query independently. Furthermore, we also note that TSAspire and OTAspire by their design can only encode both the query and the abstract into multiple vectors, as they utilize contextualized sentence embedding vectors for "aspects". Note, they do not process each sentence independently. These two models employ the distance function described in \cite{aspire} to assess the relationship between the query vectors and the abstract vectors. 

To calculate relevance between the query $Q$, which is encoded into a list of vectors $\{\vec{Q_1},\dots, \vec{Q_n}\}$ and the abstract $p$, which is encoded into a list of vectors $\{\vec{p_1},\dots, \vec{p_m}\}$, we employ the standard max-sim operation used in \cite{colbertv1}, specified by Equation \ref{eq:maxsim}. We report our results in Table \ref{tab:ranking_query_level_sent_by_sent_appendix}.

\begin{equation}
    \label{eq:maxsim}
    \textrm{Rel}(p|Q) = \frac{1}{n} \sum_{i=1}^n \max_{1\leq j\leq m} \frac{<\vec{Q_i},\vec{p_j}>}{||\vec{Q_i}||\cdot ||\vec{p_j}||}  
\end{equation}


We can see RocketQA-v2 benefited the most from processing the text sentence by sentence, reaching the highest recorded performance across 6 metrics so far, which suggests a potential reason for the weakness of other NIR models, that they are unsuitable to process longer and more complex queries.

\textcolor{black}{As mentioned before, in Table \ref{tab:query-level ranking} in the main paper, we took the best recorded metrics for each model over these two options (Table \ref{tab: Query-level read all appendix} and Table \ref{tab:ranking_query_level_sent_by_sent_appendix}). }


\begin{table}[ht]
\centering
\footnotesize
\caption{Models' query-level ranking performance (embedding query and abstract into multiple vectors)}
\vspace{1em}
\setlength{\tabcolsep}{3pt}
\begin{tabular}{lcccccccc}
\toprule
Method & R@5 & R@20 & RP & $\textrm{NDCG}\textsubscript{10\%}$ & $\textrm{NDCG}^{\textrm{exp}}_{10\%}$ & MRR@10 & MAP \\
\midrule
random & 4.41 & 18.48 & 16.29 & 48.72 & 7.31 & 3.59 & 19.63 \\
\midrule
SentBERT & 14.16 \scriptsize{$\pm$ 1.95} & 44.81 \scriptsize{$\pm$ 2.54} & 33.84 \scriptsize{$\pm$ 2.50} & 69.58 \scriptsize{$\pm$ 0.96} & 22.03 \scriptsize{$\pm$ 2.06} & 13.39 \scriptsize{$\pm$ 2.63} & 37.89 \scriptsize{$\pm$ 2.34} \\
RocketQA & \textbf{15.55} \scriptsize{$\pm$ 1.93} & \textbf{45.38 }\scriptsize{$\pm$ 2.43} & \textbf{34.43} \scriptsize{$\pm$ 2.39} & \textbf{73.07} \scriptsize{$\pm$ 1.02}& \textbf{30.27} \scriptsize{$\pm$ 2.29} & \textbf{20.79} \scriptsize{$\pm$ 3.06} & \textbf{40.25} \scriptsize{$\pm$ 2.28} \\
SPLADE-v2 & 12.84 \scriptsize{$\pm$ 1.83} & 37.86 \scriptsize{$\pm$ 2.41} & 29.30 \scriptsize{$\pm$ 2.40} & 68.06 \scriptsize{$\pm$ 0.98} & 23.78 \scriptsize{$\pm$ 2.13} & 16.53 \scriptsize{$\pm$ 2.71} & 34.67 \scriptsize{$\pm$ 2.16} \\

TSAspire & 14.26 \scriptsize{$\pm$ 1.80} & 41.25 \scriptsize{$\pm$ 2.40} & 33.81 \scriptsize{$\pm$ 2.47} & 71.63 \scriptsize{$\pm$ 1.11} & 26.63 \scriptsize{$\pm$ 2.05} & 15.59 \scriptsize{$\pm$ 2.59} & 37.00 \scriptsize{$\pm$ 2.29} \\
OTAspire & 13.34 \scriptsize{$\pm$ 1.56} & 42.25 \scriptsize{$\pm$ 2.53} & 33.63 \scriptsize{$\pm$ 2.38} & 70.43 \scriptsize{$\pm$ 1.08} & 25.52 \scriptsize{$\pm$ 2.29} & 14.18 \scriptsize{$\pm$ 2.66} & 36.70 \scriptsize{$\pm$ 2.22} \\
ANCE & 13.17 \scriptsize{$\pm$ 1.89} & 34.55 \scriptsize{$\pm$ 2.11} & 30.33 \scriptsize{$\pm$ 2.50} & 66.83 \scriptsize{$\pm$ 1.01} & 20.27 \scriptsize{$\pm$ 1.91} & 13.83 \scriptsize{$\pm$ 2.45} & 34.37 \scriptsize{$\pm$ 2.33} \\
SimLM & 13.77 \scriptsize{$\pm$ 1.87} & 38.11 \scriptsize{$\pm$ 2.26} & 30.17 \scriptsize{$\pm$ 2.30} & 67.35 \scriptsize{$\pm$ 1.14} & 23.61 \scriptsize{$\pm$ 2.14} & 16.63 \scriptsize{$\pm$ 2.88} & 34.94 \scriptsize{$\pm$ 2.14} \\


ColBERT-v2 & 9.19 \scriptsize{$\pm$ 1.27} & 30.28 \scriptsize{$\pm$ 2.03} & 23.77 \scriptsize{$\pm$ 1.97} & 61.99 \scriptsize{$\pm$ 1.05} & 13.79 \scriptsize{$\pm$ 1.48} & 7.20 \scriptsize{$\pm$ 2.13} & 28.36 \scriptsize{$\pm$ 1.84} \\

SciBERT & 5.37 \scriptsize{$\pm$ 0.83} & 22.34 \scriptsize{$\pm$ 2.01} & 19.44 \scriptsize{$\pm$ 1.86} & 53.19 \scriptsize{$\pm$ 1.24} & 8.01 \scriptsize{$\pm$ 1.17} & 4.08 \scriptsize{$\pm$ 1.38} & 22.54 \scriptsize{$\pm$ 1.62} \\

SciBERT\textsubscript{ID} & 12.46 \scriptsize{$\pm$ 1.88} & 35.40 \scriptsize{$\pm$ 2.17} & 27.24 \scriptsize{$\pm$ 2.27} & 64.67 \scriptsize{$\pm$ 1.18} & 18.42 \scriptsize{$\pm$ 2.08} & 12.74 \scriptsize{$\pm$ 2.63} & 32.01 \scriptsize{$\pm$ 2.03} \\
ERNIE & 6.71 \scriptsize{$\pm$ 1.39} & 24.24 \scriptsize{$\pm$ 1.96} & 21.39 \scriptsize{$\pm$ 2.08} & 54.27 \scriptsize{$\pm$ 1.02} & 9.92 \scriptsize{$\pm$ 1.50} & 6.05 \scriptsize{$\pm$ 2.02} & 24.00 \scriptsize{$\pm$ 1.81} \\
SimCSE & 13.21 \scriptsize{$\pm$ 1.86} & 35.83 \scriptsize{$\pm$ 2.31} & 29.41 \scriptsize{$\pm$ 2.23} & 65.20 \scriptsize{$\pm$ 1.17} & 20.62 \scriptsize{$\pm$ 1.84} & 13.59 \scriptsize{$\pm$ 2.58} & 33.89 \scriptsize{$\pm$ 2.17} \\

\bottomrule
\end{tabular}
\label{tab:ranking_query_level_sent_by_sent_appendix}
\end{table}

\subsection{Using Aspects for Retrieval}
\label{ssec:Aspects are visible}

In Section \ref{sec: additional_experiments}, we explored using the hierarchical structure of aspects and sub-aspects as an alternative to traditional queries. For this analysis, we concatenated the aspects into a paragraph to compute query embeddings for various models. However, this resulted in a verbose and repetitive paragraph, which did not resemble a real query. Given that some methods are trained specifically for sentence-level tasks, we offered the option of using a multi-vector representation for the aspects, with one vector per aspect. \textcolor{black}{Similar to Sections \ref{ssec:read the whole query once} and \ref{ssec: read sentence by sentence}, we considered two ways to process queries and abstracts: Approach 1 concatenates aspects and computes a single embedding; Approach 2 embeds aspects and abstracts sentence-by-sentence (this is denoted with a * in the table). For clarity, we only record the best metrics for each model between these two options in Table \ref{tab:ranking_aspect_appendix}.}

When we compared Table \ref{tab:ranking_aspect_appendix} with Tables \ref{tab:ranking_query_level_sent_by_sent_appendix} and \ref{tab: Query-level read all appendix}, we did not observe any clear effect of using aspects for retrieval. Sentence-BERT and SPLADE-v2 did achieve stronger results in this setting, but the effect did not generalize to other models. 

\begin{table}[ht]
\centering
\footnotesize
\caption{\textcolor{black}{Models' ranking performance using concatenated aspects as query}}
\vspace{1em}
\setlength{\tabcolsep}{3pt}
\begin{tabular}{lcccccccc}
\toprule
Method & R@5 & R@20 & RP & $\textrm{NDCG}\textsubscript{10\%}$ & $\textrm{NDCG}^{\textrm{exp}}_{10\%}$ & MRR@10 & MAP \\
\midrule
random & 4.60 & 18.30 & 16.48 & 48.79 & 7.41 & 3.65 & 19.73 \\
\midrule
SentBERT* & \textbf{17.71} \scriptsize{$\pm$ 2.44} & \textbf{45.26} \scriptsize{$\pm$ 2.66} & \textbf{35.63} \scriptsize{$\pm$ 2.48} & 70.47 \scriptsize{$\pm$ 0.97} & 24.94 \scriptsize{$\pm$ 2.19} & 15.54 \scriptsize{$\pm$ 3.00} & \textbf{39.79} \scriptsize{$\pm$ 2.42} \\
RocketQA* & 13.79 \scriptsize{$\pm$ 1.48} & 43.69 \scriptsize{$\pm$ 2.38} & 32.63 \scriptsize{$\pm$ 2.17} & \textbf{72.21} \scriptsize{$\pm$ 0.93}& \textbf{27.39} \scriptsize{$\pm$ 2.04} & 16.03 \scriptsize{$\pm$ 2.52} & 37.90 \scriptsize{$\pm$ 2.13} \\
ANCE* & 15.35 \scriptsize{$\pm$ 2.02} & 38.36 \scriptsize{$\pm$ 2.29} & 29.45 \scriptsize{$\pm$ 2.07} & 67.27 \scriptsize{$\pm$ 1.08} & 20.46 \scriptsize{$\pm$ 1.82} & 14.54 \scriptsize{$\pm$ 2.75} & 33.59 \scriptsize{$\pm$ 1.89} \\
SPLADE-v2 & 15.07 \scriptsize{$\pm$ 1.99} & 41.20 \scriptsize{$\pm$ 2.42} & 34.26 \scriptsize{$\pm$ 2.45} & 71.34 \scriptsize{$\pm$ 0.91} & 24.91 \scriptsize{$\pm$ 1.90} & 18.30 \scriptsize{$\pm$ 3.00} & 38.72 \scriptsize{$\pm$ 2.33} \\
ColBERT-v2* & 7.37 \scriptsize{$\pm$ 1.11} & 26.64 \scriptsize{$\pm$ 1.94} & 22.28 \scriptsize{$\pm$ 1.82} & 58.70 \scriptsize{$\pm$ 1.14} & 12.17 \scriptsize{$\pm$ 1.30} & 5.78 \scriptsize{$\pm$ 1.65} & 25.24 \scriptsize{$\pm$ 1.69} \\
ERNIE* & 5.34 \scriptsize{$\pm$ 1.23} & 19.96 \scriptsize{$\pm$ 1.58} & 18.02 \scriptsize{$\pm$ 1.64} & 50.33 \scriptsize{$\pm$ 1.13} & 7.14 \scriptsize{$\pm$ 1.03} & 3.14 \scriptsize{$\pm$ 1.26} & 21.38 \scriptsize{$\pm$ 1.50} \\
SciBERT* & 5.41 \scriptsize{$\pm$ 0.78} & 21.36 \scriptsize{$\pm$ 1.86} & 18.50 \scriptsize{$\pm$ 1.72} & 51.27 \scriptsize{$\pm$ 1.15} & 6.84 \scriptsize{$\pm$ 0.92} & 2.29 \scriptsize{$\pm$ 1.04} & 21.66 \scriptsize{$\pm$ 1.57} \\
SciBERT\textsubscript{ID}* & 8.88 \scriptsize{$\pm$ 1.44} & 30.81 \scriptsize{$\pm$ 2.15} & 22.56 \scriptsize{$\pm$ 2.00} & 59.86 \scriptsize{$\pm$ 1.01} & 13.82 \scriptsize{$\pm$ 1.64} & 10.24 \scriptsize{$\pm$ 2.28} & 26.53 \scriptsize{$\pm$ 1.77} \\
ada-002 & 14.12 \scriptsize{$\pm$ 2.00} & 42.13 \scriptsize{$\pm$ 2.73} & 33.60 \scriptsize{$\pm$ 2.44} & 71.52 \scriptsize{$\pm$ 0.98} & 26.57 \scriptsize{$\pm$ 2.57} & \textbf{20.22} \scriptsize{$\pm$ 3.49} & 37.63 \scriptsize{$\pm$ 2.29} \\
E5-L-v2 & 14.32 \scriptsize{$\pm$ 1.97} & 41.55 \scriptsize{$\pm$ 2.21} & 34.21 \scriptsize{$\pm$ 2.54} & 69.37 \scriptsize{$\pm$ 1.00} & 22.51 \scriptsize{$\pm$ 2.16} & 13.87 \scriptsize{$\pm$ 2.68} & 37.31 \scriptsize{$\pm$ 2.33} \\
TSAspire & 14.30 \scriptsize{$\pm$ 1.80} & 42.75 \scriptsize{$\pm$ 2.25} & 33.63 \scriptsize{$\pm$ 2.45} & 70.96 \scriptsize{$\pm$ 1.10} & 25.29 \scriptsize{$\pm$ 2.14} & 16.26 \scriptsize{$\pm$ 2.90} & 37.31 \scriptsize{$\pm$ 2.23} \\
OTAspire & 14.45 \scriptsize{$\pm$ 1.81} & 40.99 \scriptsize{$\pm$ 2.30} & 32.53 \scriptsize{$\pm$ 2.43} & 69.74 \scriptsize{$\pm$ 1.14} & 23.81 \scriptsize{$\pm$ 2.07} & 15.38 \scriptsize{$\pm$ 2.76} & 36.09 \scriptsize{$\pm$ 2.18} \\
SimLM & 16.18 \scriptsize{$\pm$ 2.19} & 39.24 \scriptsize{$\pm$ 2.36} & 32.20 \scriptsize{$\pm$ 2.22} & 67.56 \scriptsize{$\pm$ 1.03} & 23.36 \scriptsize{$\pm$ 2.14} & 16.61 \scriptsize{$\pm$ 2.84} & 36.24 \scriptsize{$\pm$ 2.12} \\
SimCSE & 12.41 \scriptsize{$\pm$ 1.84} & 35.02 \scriptsize{$\pm$ 2.52} & 27.67 \scriptsize{$\pm$ 2.27} & 64.19 \scriptsize{$\pm$ 1.24} & 20.52 \scriptsize{$\pm$ 1.93} & 15.27 \scriptsize{$\pm$ 2.57} & 30.83 \scriptsize{$\pm$ 2.12} \\

SPECTER & 11.56 \scriptsize{$\pm$ 1.76} & 35.30 \scriptsize{$\pm$ 2.34} & 28.43 \scriptsize{$\pm$ 2.49} & 63.55 \scriptsize{$\pm$ 1.30} & 19.69 \scriptsize{$\pm$ 2.33} & 15.07 \scriptsize{$\pm$ 3.12} & 31.39 \scriptsize{$\pm$ 2.22} \\
SPECTER\textsubscript{ID} & 12.01 \scriptsize{$\pm$ 1.77} & 35.79 \scriptsize{$\pm$ 2.27} & 28.71 \scriptsize{$\pm$ 2.17} & 66.22 \scriptsize{$\pm$ 1.05} & 19.70 \scriptsize{$\pm$ 2.04} & 13.44 \scriptsize{$\pm$ 2.67} & 32.69 \scriptsize{$\pm$ 1.97} \\
SPECTER-v2 & 12.97 \scriptsize{$\pm$ 1.78} & 41.45 \scriptsize{$\pm$ 2.50} & 31.30 \scriptsize{$\pm$ 2.35} & 68.37 \scriptsize{$\pm$ 1.13} & 23.16 \scriptsize{$\pm$ 2.20} & 14.59 \scriptsize{$\pm$ 2.66} & 35.77 \scriptsize{$\pm$ 2.09} \\

BM25 & 6.25 \scriptsize{$\pm$ 1.00} & 25.11 \scriptsize{$\pm$ 2.09} & 20.51 \scriptsize{$\pm$ 1.88} & 49.65 \scriptsize{$\pm$ 1.31} & 10.33 \scriptsize{$\pm$ 1.33} & 5.72 \scriptsize{$\pm$ 1.62} & 23.74 \scriptsize{$\pm$ 1.57} \\
TF-IDF & 7.62 \scriptsize{$\pm$ 1.40} & 23.16 \scriptsize{$\pm$ 2.10} & 20.14 \scriptsize{$\pm$ 2.09} & 50.73 \scriptsize{$\pm$ 1.52} & 12.55 \scriptsize{$\pm$ 1.81} & 7.72 \scriptsize{$\pm$ 2.15} & 24.25 \scriptsize{$\pm$ 1.91} \\

\bottomrule
\multicolumn{8}{l}{{ * : Approach 2}} \\
\end{tabular}
\label{tab:ranking_aspect_appendix}
\end{table}



\subsection{Model Performance on DORIS-MAE Test Set}
\label{ssec:test set}
In Appendix section \ref{ssec:query level benchmark results}, we evaluated zero-shot performance of embedding models on the full DORIS-MAE dataset. We now present zero-shot results on the DORIS-MAE test set, which consists of 60 queries. For these queries, we run all IR models and show their performances in Table \ref{tab: 60 query}. All models use their individual optimal embedding method (i.e. processing query/abstract sentence by sentence or as an entire paragraph). Results from this table serve as a baseline for future work that trains models on the DORIS-MAE training set. 

\begin{table}[ht]
\centering
\footnotesize
\caption{Models' query-level ranking performance on DORIS-MAE test set}
\vspace{1em}
\setlength{\tabcolsep}{3pt}
\begin{tabular}{lcccccccc}
\toprule
Method & R@5 & R@20 & RP & $\textrm{NDCG}\textsubscript{10\%}$ & $\textrm{NDCG}^{\textrm{exp}}_{10\%}$ & MRR@10 & MAP \\
\midrule
random & 4.58 & 17.91 & 17.34 & 50.22 & 7.36  & 3.32 & 20.69\\
\midrule
ada-002 & \textbf{15.80} \scriptsize{$\pm$ 1.92} & 40.87 \scriptsize{$\pm$ 3.09} & 37.64 \scriptsize{$\pm$ 3.17} & \textbf{73.92} \scriptsize{$\pm$ 1.29} & 28.47 \scriptsize{$\pm$ 3.22} & 23.29 \scriptsize{$\pm$ 4.52} & \textbf{41.72} \scriptsize{$\pm$ 3.12} \\
SimCSE & 14.40 \scriptsize{$\pm$ 1.71} & 41.94 \scriptsize{$\pm$ 3.01} & 36.29 \scriptsize{$\pm$ 2.86} & 71.31 \scriptsize{$\pm$ 1.25} & 29.08 \scriptsize{$\pm$ 3.16} & \textbf{25.00} \scriptsize{$\pm$ 4.85} & 39.65 \scriptsize{$\pm$ 2.86} \\
RocketQA & 13.26 \scriptsize{$\pm$ 1.29} & \textbf{44.35} \scriptsize{$\pm$ 2.80} & 35.32 \scriptsize{$\pm$ 2.97} & 73.87 \scriptsize{$\pm$ 1.26} & \textbf{29.83} \scriptsize{$\pm$ 2.88} & 20.61 \scriptsize{$\pm$ 4.01} & 40.66 \scriptsize{$\pm$ 2.84} \\
SPECTER-v2 & 13.33 \scriptsize{$\pm$ 2.26} & 41.83 \scriptsize{$\pm$ 3.07} & 35.98 \scriptsize{$\pm$ 3.16} & 71.46 \scriptsize{$\pm$ 1.15} & 27.67 \scriptsize{$\pm$ 3.26} & 20.70 \scriptsize{$\pm$ 4.47} & 38.46 \scriptsize{$\pm$ 2.80} \\
TSAspire & 13.31 \scriptsize{$\pm$ 1.67} & 43.57 \scriptsize{$\pm$ 2.97} & 36.51 \scriptsize{$\pm$ 2.96} & 72.78 \scriptsize{$\pm$ 1.26} & 27.72 \scriptsize{$\pm$ 2.81} & 18.22 \scriptsize{$\pm$ 3.81} & 38.50 \scriptsize{$\pm$ 2.87} \\
E5-L-v2 & 14.59 \scriptsize{$\pm$ 1.79} & 42.06 \scriptsize{$\pm$ 2.47} & \textbf{38.16} \scriptsize{$\pm$ 2.93} & 71.96 \scriptsize{$\pm$ 1.18} & 26.04 \scriptsize{$\pm$ 2.76} & 14.25 \scriptsize{$\pm$ 3.52} & 40.51 \scriptsize{$\pm$ 2.87} \\
OTAspire & 13.12 \scriptsize{$\pm$ 1.61} & 42.65 \scriptsize{$\pm$ 3.22} & 35.35 \scriptsize{$\pm$ 2.97} & 71.01 \scriptsize{$\pm$ 1.36} & 25.69 \scriptsize{$\pm$ 2.72} & 15.38 \scriptsize{$\pm$ 3.22} & 37.83 \scriptsize{$\pm$ 2.86} \\
SPLADE-v2 & 10.99 \scriptsize{$\pm$ 1.18} & 36.21 \scriptsize{$\pm$ 2.57} &  32.50 \scriptsize{$\pm$ 2.96} & 71.01 \scriptsize{$\pm$  1.18} & 25.10 \scriptsize{$\pm$  2.90} & 19.44 \scriptsize{$\pm$ 4.38} & 36.44 \scriptsize{$\pm$ 2.81} \\
SentBERT & 12.18 \scriptsize{$\pm$ 1.72} & 43.23 \scriptsize{$\pm$ 2.98} & 34.38 \scriptsize{$\pm$ 3.07} & 69.63 \scriptsize{$\pm$ 1.31} & 19.37 \scriptsize{$\pm$ 2.47} & 10.84 \scriptsize{$\pm$ 3.34} & 37.72 \scriptsize{$\pm$ 2.93} \\
SimLM & 10.49 \scriptsize{$\pm$ 1.27} & 32.19 \scriptsize{$\pm$ 2.72} & 29.73 \scriptsize{$\pm$ 3.14} & 66.73 \scriptsize{$\pm$ 1.36} & 17.50 \scriptsize{$\pm$ 2.52} & 11.90 \scriptsize{$\pm$ 3.66} & 32.28 \scriptsize{$\pm$ 3.04} \\
ANCE\textsubscript{FirstP} & 10.88 \scriptsize{$\pm$ 1.83} & 32.03 \scriptsize{$\pm$ 2.72} & 29.02 \scriptsize{$\pm$ 3.04} & 66.48 \scriptsize{$\pm$ 1.32} & 17.92 \scriptsize{$\pm$ 2.28} & 11.11 \scriptsize{$\pm$ 2.79} & 32.90 \scriptsize{$\pm$ 2.83} \\
LLAMA & 9.60 \scriptsize{$\pm$ 1.36} & 32.34 \scriptsize{$\pm$ 2.86} & 28.43 \scriptsize{$\pm$ 2.69} & 65.15 \scriptsize{$\pm$ 1.24} & 14.71 \scriptsize{$\pm$ 1.84} & 9.36 \scriptsize{$\pm$ 2.67} & 31.16 \scriptsize{$\pm$ 2.60} \\
TF-IDF & 9.53 \scriptsize{$\pm$ 1.42} & 26.77 \scriptsize{$\pm$ 2.51} & 25.27 \scriptsize{$\pm$ 2.50} & 58.14 \scriptsize{$\pm$ 1.59} & 17.03 \scriptsize{$\pm$ 2.39} & 11.59 \scriptsize{$\pm$ 3.36} & 28.56 \scriptsize{$\pm$ 2.28} \\
BM25 & 6.86 \scriptsize{$\pm$ 1.52} & 26.43 \scriptsize{$\pm$ 2.27} & 22.45 \scriptsize{$\pm$ 2.51} & 51.55 \scriptsize{$\pm$ 1.76} & 12.29 \scriptsize{$\pm$ 2.55} & 8.65 \scriptsize{$\pm$ 3.49} & 26.00 \scriptsize{$\pm$ 2.16} \\
ColBERTv2 & 6.49 \scriptsize{$\pm$ 1.21} & 23.99 \scriptsize{$\pm$ 2.78} & 21.58 \scriptsize{$\pm$ 2.66} & 59.46 \scriptsize{$\pm$ 1.64} & 12.92 \scriptsize{$\pm$ 2.49} & 7.15 \scriptsize{$\pm$ 2.97} & 25.24 \scriptsize{$\pm$ 2.41} \\
ERNIE & 7.43 \scriptsize{$\pm$ 1.26} & 24.92 \scriptsize{$\pm$ 2.23} & 22.43 \scriptsize{$\pm$ 2.50} & 57.52 \scriptsize{$\pm$ 1.50} & 10.83 \scriptsize{$\pm$ 1.62} & 3.64 \scriptsize{$\pm$ 1.28} & 24.94 \scriptsize{$\pm$ 2.24} \\
SciBERT & 4.57 \scriptsize{$\pm$ 1.08} & 17.35 \scriptsize{$\pm$ 1.84} & 16.83 \scriptsize{$\pm$ 2.27} & 49.42 \scriptsize{$\pm$ 1.56} & 7.13 \scriptsize{$\pm$ 1.19} & 2.00 \scriptsize{$\pm$ 1.06} & 20.61 \scriptsize{$\pm$ 1.91} \\

\bottomrule
\end{tabular}
\label{tab: 60 query}
\end{table}

\subsection{Two-Aspect Sub-Query Level Benchmark Results}
\label{ssec:sub-query level benchmark results}

The dataset's hierarchical structure provides an opportunity to formulate simpler tasks involving only parts of a complex query. By extracting parts corresponding to two aspects, we generated over \textcolor{black}{1000} test cases, significantly increasing the number of relevant abstracts as outlined in Section \ref{ssec:statistics of dataset}. The full results for this task are given in Table \ref{tab:ranking_subquery_appendix}. As the length of these sub-queries is similar to a long sentence, we did not differentiate between embedding the entire query as a single vector or multiple vectors sentence-by-sentence. Except for OTAspire and TSAspire, we applied a single-vector embedding for all other models, including Sentence-BERT and ANCE.

\textcolor{black}{We observed that ada-002 outperformed other models across all categories. All models exhibited slight increases in R-Precision, $\textrm{NDCG}^{\textrm{exp}}_{10\%}$, and MAP. These minor improvements correlate with similar increases in their random baselines, suggesting that they might be due to shifts in the random baseline rather than the models' retrieval capabilities.}\textcolor{black}{ Meanwhile, the small decreases in the random baselines of R@5 and R@20 indicate a drop in the retrieval models' Recall@k performance on the sub-query level. This corresponds to an increase in the number of relevant abstracts per sub-query as shown in Table \ref{tab:statistics_of_relevant_abstracts}. As Recall@k is inversely related to the number of relevant abstracts per sub-query, and as the retrieval models do not appear to effectively retrieve relevant abstracts for these simpler sub-queries, it is unsurprising that their R@5 and R@20 metrics dropped substantially.}

\begin{table}[ht]
\centering
\footnotesize
\caption{\textcolor{black}{Models' ranking performance on 2-aspect sub-queries}}
\vspace{1em}
\setlength{\tabcolsep}{3pt}
\begin{tabular}{lcccccccc}
\toprule
Method & R@5 & R@20 & RP & $\textrm{NDCG}\textsubscript{10\%}$ & $\textrm{NDCG}^{\textrm{exp}}_{10\%}$ & MRR@10 & MAP \\
\midrule
random & 4.60 & 18.76 & 21.55 & 41.10 & 15.06 & 5.92 & 24.71 \\
\midrule
ada-002 & \textbf{13.47} \scriptsize{$\pm$ 0.47} & 40.22 \scriptsize{$\pm$ 0.64} & 47.34 \scriptsize{$\pm$ 0.77} & \textbf{69.78} \scriptsize{$\pm$ 0.39} & \textbf{38.97} \scriptsize{$\pm$ 0.64} & \textbf{24.29} \scriptsize{$\pm$ 1.07} & \textbf{51.65} \scriptsize{$\pm$ 0.77} \\
LLAMA & 8.06 \scriptsize{$\pm$ 0.31} & 29.77 \scriptsize{$\pm$ 0.61} & 36.49 \scriptsize{$\pm$ 0.69} & 56.85 \scriptsize{$\pm$ 0.44} & 23.59 \scriptsize{$\pm$ 0.50} & 9.80 \scriptsize{$\pm$ 0.66} & 39.01 \scriptsize{$\pm$ 0.65} \\
E5-L-v2 & 11.62 \scriptsize{$\pm$ 0.37} & 38.26 \scriptsize{$\pm$ 0.58} & 45.11 \scriptsize{$\pm$ 0.78} & 65.82 \scriptsize{$\pm$ 0.42} & 35.12 \scriptsize{$\pm$ 0.59} & 21.00 \scriptsize{$\pm$ 0.95} & 48.73 \scriptsize{$\pm$ 0.76} \\
TSAspire & 13.15 \scriptsize{$\pm$ 0.69} & 43.47 \scriptsize{$\pm$ 1.04} & 41.46 \scriptsize{$\pm$ 1.24} & 65.54 \scriptsize{$\pm$ 0.71} & 36.31 \scriptsize{$\pm$ 0.97} & 23.54 \scriptsize{$\pm$ 1.58} & 45.85 \scriptsize{$\pm$ 1.20} \\
OTAspire & 12.56 \scriptsize{$\pm$ 0.70} & \textbf{43.66} \scriptsize{$\pm$ 1.09} & 40.60 \scriptsize{$\pm$ 1.23} & 65.06 \scriptsize{$\pm$ 0.69} & 35.54 \scriptsize{$\pm$ 0.91} & 22.71 \scriptsize{$\pm$ 1.49} & 45.53 \scriptsize{$\pm$ 1.16} \\
SPECTER & 10.51 \scriptsize{$\pm$ 0.85} & 35.99 \scriptsize{$\pm$ 1.07} & 35.40 \scriptsize{$\pm$ 1.13} & 55.71 \scriptsize{$\pm$ 0.67} & 26.26 \scriptsize{$\pm$ 0.95} & 14.06 \scriptsize{$\pm$ 1.32} & 38.27 \scriptsize{$\pm$ 1.05} \\
SPECTER-v2 & 11.14 \scriptsize{$\pm$ 0.41} & 36.09 \scriptsize{$\pm$ 0.60} & 43.73 \scriptsize{$\pm$ 0.80} & 64.80 \scriptsize{$\pm$ 0.41} & 32.68 \scriptsize{$\pm$ 0.55} & 18.02 \scriptsize{$\pm$ 0.90} & 47.26 \scriptsize{$\pm$ 0.79} \\
SPECTER\textsubscript{ID} & 8.05 \scriptsize{$\pm$ 0.31} & 30.97 \scriptsize{$\pm$ 0.56} & 38.11 \scriptsize{$\pm$ 0.77} & 57.19 \scriptsize{$\pm$ 0.43} & 23.96 \scriptsize{$\pm$ 0.45} & 10.09 \scriptsize{$\pm$ 0.64} & 40.23 \scriptsize{$\pm$ 0.74} \\
SentBERT & 12.79 \scriptsize{$\pm$ 0.42} & 40.31 \scriptsize{$\pm$ 0.62} & \textbf{48.34} \scriptsize{$\pm$ 0.80} & 69.38 \scriptsize{$\pm$ 0.42} & 38.52 \scriptsize{$\pm$ 0.61} & 22.98 \scriptsize{$\pm$ 1.00} & 52.34 \scriptsize{$\pm$ 0.80} \\
RocketQA & 13.37 \scriptsize{$\pm$ 0.61} & 42.42 \scriptsize{$\pm$ 1.00} & 42.65 \scriptsize{$\pm$ 1.19} & 67.20 \scriptsize{$\pm$ 0.70} & 37.33 \scriptsize{$\pm$ 1.04} & 24.19 \scriptsize{$\pm$ 1.65} & 46.39 \scriptsize{$\pm$ 1.21} \\
ANCE\textsubscript{FirstP} & 11.23 \scriptsize{$\pm$ 0.43} & 35.12 \scriptsize{$\pm$ 0.58} & 42.49 \scriptsize{$\pm$ 0.77} & 64.00 \scriptsize{$\pm$ 0.43} & 31.64 \scriptsize{$\pm$ 0.62} & 17.70 \scriptsize{$\pm$ 0.96} & 45.78 \scriptsize{$\pm$ 0.77} \\
SimLM & 12.61 \scriptsize{$\pm$ 0.70} & 37.73 \scriptsize{$\pm$ 1.02} & 37.90 \scriptsize{$\pm$ 1.08} & 63.18 \scriptsize{$\pm$ 0.69} & 31.59 \scriptsize{$\pm$ 0.95} & 19.18 \scriptsize{$\pm$ 1.46} & 41.82 \scriptsize{$\pm$ 1.10} \\
SPLADE-v2 & 12.95 \scriptsize{$\pm$ 0.46} & 39.55 \scriptsize{$\pm$ 0.62} & 46.63 \scriptsize{$\pm$ 0.75} & 68.26 \scriptsize{$\pm$ 0.39} & 37.33 \scriptsize{$\pm$ 0.60} & 22.80 \scriptsize{$\pm$ 1.00} & 50.71 \scriptsize{$\pm$ 0.75} \\
ColBERT-v2 & 8.13 \scriptsize{$\pm$ 0.50} & 27.22 \scriptsize{$\pm$ 0.87} & 28.38 \scriptsize{$\pm$ 1.02} & 53.32 \scriptsize{$\pm$ 0.71} & 22.72 \scriptsize{$\pm$ 0.79} & 11.84 \scriptsize{$\pm$ 1.19} & 32.03 \scriptsize{$\pm$ 0.98} \\
ERNIE & 3.25 \scriptsize{$\pm$ 0.21} & 15.96 \scriptsize{$\pm$ 0.53} & 20.05 \scriptsize{$\pm$ 0.87} & 39.26 \scriptsize{$\pm$ 0.69} & 12.83 \scriptsize{$\pm$ 0.56} & 3.29 \scriptsize{$\pm$ 0.66} & 23.27 \scriptsize{$\pm$ 0.79} \\
SciBERT & 2.43 \scriptsize{$\pm$ 0.20} & 14.59 \scriptsize{$\pm$ 0.71} & 17.56 \scriptsize{$\pm$ 0.78} & 33.29 \scriptsize{$\pm$ 0.58} & 9.65 \scriptsize{$\pm$ 0.42} & 1.35 \scriptsize{$\pm$ 0.24} & 21.23 \scriptsize{$\pm$ 0.72} \\
SciBERT\textsubscript{ID} & 9.41 \scriptsize{$\pm$ 0.35} & 32.43 \scriptsize{$\pm$ 0.56} & 39.28 \scriptsize{$\pm$ 0.72} & 60.19 \scriptsize{$\pm$ 0.42} & 27.91 \scriptsize{$\pm$ 0.51} & 13.63 \scriptsize{$\pm$ 0.80} & 42.19 \scriptsize{$\pm$ 0.70} \\
BM25 & 9.14 \scriptsize{$\pm$ 0.62} & 31.68 \scriptsize{$\pm$ 0.99} & 31.22 \scriptsize{$\pm$ 1.00} & 51.05 \scriptsize{$\pm$ 0.71} & 25.52 \scriptsize{$\pm$ 0.84} & 14.74 \scriptsize{$\pm$ 1.18} & 33.94 \scriptsize{$\pm$ 0.94} \\
TF-IDF & 9.69 \scriptsize{$\pm$ 0.58} & 32.56 \scriptsize{$\pm$ 0.93} & 33.32 \scriptsize{$\pm$ 1.05} & 56.58 \scriptsize{$\pm$ 0.67} & 28.79 \scriptsize{$\pm$ 0.91} & 19.57 \scriptsize{$\pm$ 1.52} & 36.41 \scriptsize{$\pm$ 0.97} \\
SimCSE & 10.98 \scriptsize{$\pm$ 0.38} & 34.57 \scriptsize{$\pm$ 0.58} & 41.40 \scriptsize{$\pm$ 0.74} & 63.00 \scriptsize{$\pm$ 0.43} & 31.19 \scriptsize{$\pm$ 0.56} & 17.22 \scriptsize{$\pm$ 0.90} & 44.96 \scriptsize{$\pm$ 0.74} \\
\bottomrule
\end{tabular}
\label{tab:ranking_subquery_appendix}
\end{table}

\subsection{Candidate Pool Sensitivity Analyses}
\label{ssec:extending candidate pools}

In this section, we perform a sensitivity analysis on the candidate pools in DORIS-MAE. The original candidate pools were primarily constructed using keyword-based methods. To evaluate how this may have influenced our results, we generated a second set of candidate pools for the first 50 queries in DORIS-MAE. These new "embedding-based candidate pools" were generated with two text-embedding models, E5-v2 and SPECTER-v2, which were not used during the original candidate pool creation. 

For each query, we collect the top-ranking 150 abstracts that are not in the query's original candidate pool (75 from E5-v2 and 75 from SPECTER-v2). We applied the same LLM annotation procedure to estimate the relevance of each abstract in the candidate pool given the query.  

\textbf{Relevance Analysis:} We first evaluated whether the new pool contains documents which are more relevant than those in the original pool. For each of the 50 queries, and for each candidate pool (i.e. the original pool and the extended embedding pool), we record the highest relevance score. For each query, we then compare the highest relevance score from each pooling method. On average, the most relevant documents in the original pool scored higher than those in the embedding pool by 4\%. However, for 30\% of the queries, the embedding pool did contain a more relevant document, with an average improvement of 12\% over the original pool. The results indicate that the best documents in the original candidate pool were comparable to those in the extended embedding pool.


\textbf{Reranking Performance Analysis:} Next, we compared the reranking performance of the IR models on the two pools. Comparing Table \ref{tab: 150 candidate} with Table \ref{tab:query-level ranking} and Table \ref{tab: 60 query}, we noticed significant performance decreases in E5-v2, SPECTER-v2, SimCSE and SentBERT on the new pool. The performance of ada-002 remains stable. Since all abstracts in the new extended candidate pool are considered highly relevant by the pretrained embeddings of E5-v2 and SPECTER-v2, which are the methods used for the extended candidate pool generation, the performance drops for these two models can be attributed to selection bias that makes it harder for the same reranking models to distinguish among their own pre-selected documents. Furthermore, transformer-based encoder models may be generating similar document representations, since most of these models also see performance drops. The results suggest that it is useful to generate candidate pools using methods that have low correlation with downstream reranking methods.

\begin{table}[ht]
\centering
\footnotesize
\caption{Models' query-level ranking performance on the extended 150-abstracts candidate pool}
\vspace{1em}
\setlength{\tabcolsep}{3pt}
\begin{tabular}{lcccccccc}
\toprule
Method & R@5 & R@20 & RP & $\textrm{NDCG}\textsubscript{10\%}$ & $\textrm{NDCG}^{\textrm{exp}}_{10\%}$ & MRR@10 & MAP \\
\midrule
random & 3.28 & 13.47 & 9.36 & 47.73 & 6.71 & 3.25 & 12.37 \\
\midrule
ada-002 & \textbf{15.39} \scriptsize{$\pm$ 3.49} & \textbf{37.17} \scriptsize{$\pm$ 3.86} & \textbf{26.64} 
\scriptsize{$\pm$ 3.52} & \textbf{67.45}
\scriptsize{$\pm$ 1.42} & \textbf{23.66} \scriptsize{$\pm$ 3.34} & \textbf{20.97} \scriptsize{$\pm$ 5.20} & \textbf{29.66} \scriptsize{$\pm$ 3.35}\\
RocketQA & 11.33 \scriptsize{$\pm$ 1.70} & 36.04 \scriptsize{$\pm$ 2.57} & 21.51 
\scriptsize{$\pm$ 2.77} & 68.42
\scriptsize{$\pm$ 1.40} & 21.40 \scriptsize{$\pm$ 2.92} & 14.12 \scriptsize{$\pm$ 3.57} & 24.62 \scriptsize{$\pm$ 2.36}\\
SimCSE & 12.87 \scriptsize{$\pm$ 3.30} & 35.92 \scriptsize{$\pm$ 4.11} & 18.39 
\scriptsize{$\pm$ 2.37} & 64.22
\scriptsize{$\pm$ 1.28} & 19.24 \scriptsize{$\pm$ 1.95} & 8.34 \scriptsize{$\pm$ 2.04} & 22.16 \scriptsize{$\pm$ 1.82}\\
SentBERT & 6.55 \scriptsize{$\pm$ 1.51} & 31.21 \scriptsize{$\pm$ 3.44} & 18.41 
\scriptsize{$\pm$ 2.65} & 64.21
\scriptsize{$\pm$ 1.42} & 16.93 \scriptsize{$\pm$ 2.50} & 10.06 \scriptsize{$\pm$ 3.41} & 22.50 \scriptsize{$\pm$ 2.28}\\
TSAspire & 6.17 \scriptsize{$\pm$ 1.24} & 28.43 \scriptsize{$\pm$ 3.36} & 17.42 
\scriptsize{$\pm$ 2.36} & 62.08
\scriptsize{$\pm$ 1.20} & 13.58 \scriptsize{$\pm$ 1.74} & 4.90 \scriptsize{$\pm$ 1.39} & 19.54 \scriptsize{$\pm$ 1.92}\\
OTAspire & 6.12 \scriptsize{$\pm$ 1.21} & 26.44 \scriptsize{$\pm$ 2.95} & 17.28 
\scriptsize{$\pm$ 2.31} & 62.34
\scriptsize{$\pm$ 1.21} & 13.83 \scriptsize{$\pm$ 1.76} & 5.62 \scriptsize{$\pm$ 1.79} & 19.35 \scriptsize{$\pm$ 2.03}\\
SPLADE-v2 &  10.04 \scriptsize{$\pm$ 3.14} & 28.36 \scriptsize{$\pm$  3.45} & 14.05 
\scriptsize{$\pm$ 2.24} & 59.09
\scriptsize{$\pm$ 1.43} & 12.10 \scriptsize{$\pm$ 1.89} & 5.75 \scriptsize{$\pm$ 1.95} & 18.48 \scriptsize{$\pm$ 1.76}\\
ANCE\textsubscript{FirstP} & 9.69 \scriptsize{$\pm$ 2.61} & 24.31 \scriptsize{$\pm$ 2.77} & 17.87 
\scriptsize{$\pm$ 2.81} & 61.10
\scriptsize{$\pm$ 1.27} & 17.04 \scriptsize{$\pm$ 2.99} & 11.50 \scriptsize{$\pm$ 3.70} & 20.70 \scriptsize{$\pm$ 2.60}\\
SPECTER-v2 & 6.75 \scriptsize{$\pm$ 1.19} & 20.17 \scriptsize{$\pm$ 2.03} & 15.18 
\scriptsize{$\pm$ 1.98} & 59.11
\scriptsize{$\pm$ 1.46} & 12.54 \scriptsize{$\pm$ 1.93} & 8.45 \scriptsize{$\pm$ 3.15} & 17.37 \scriptsize{$\pm$ 1.83}\\
SimLM & 5.55 \scriptsize{$\pm$ 1.04} & 24.88 \scriptsize{$\pm$ 2.69} & 16.22 
\scriptsize{$\pm$ 2.08} & 60.54
\scriptsize{$\pm$ 1.23} & 13.27 \scriptsize{$\pm$ 1.74} & 4.76 \scriptsize{$\pm$ 2.23} & 18.11 \scriptsize{$\pm$ 1.66}\\
TF-IDF & 5.22 \scriptsize{$\pm$ 1.55} & 19.64 \scriptsize{$\pm$ 2.71} & 14.15 
\scriptsize{$\pm$ 2.24} & 55.93
\scriptsize{$\pm$ 1.38} & 15.32 \scriptsize{$\pm$ 2.97} & 9.58 \scriptsize{$\pm$ 3.59} & 17.24 \scriptsize{$\pm$ 1.90}\\
E5-L-v2 & 4.88 \scriptsize{$\pm$ 0.93} & 13.01 \scriptsize{$\pm$ 1.99} & 11.43 
\scriptsize{$\pm$ 2.00} & 48.53
\scriptsize{$\pm$ 1.80} & 9.04 \scriptsize{$\pm$ 1.59} & 2.92 \scriptsize{$\pm$ 1.24} & 14.48 \scriptsize{$\pm$ 1.90}\\
ColBERT-v2 & 6.73 \scriptsize{$\pm$ 2.35} & 17.40 \scriptsize{$\pm$ 2.58} & 11.58 
\scriptsize{$\pm$ 1.72} & 57.10
\scriptsize{$\pm$ 1.34} & 8.22 \scriptsize{$\pm$ 1.21} & 2.14 \scriptsize{$\pm$ 0.84} & 14.70 \scriptsize{$\pm$ 1.41}\\
ERNIE & 4.25 \scriptsize{$\pm$ 1.22} & 13.86 \scriptsize{$\pm$ 2.57} & 10.39 
\scriptsize{$\pm$ 1.64} & 47.05
\scriptsize{$\pm$ 1.65} & 6.91 \scriptsize{$\pm$ 1.40} & 2.78 \scriptsize{$\pm$ 1.42} & 12.89 \scriptsize{$\pm$ 1.42}\\
BM25 & 2.16 \scriptsize{$\pm$ 0.68} & 21.13 \scriptsize{$\pm$ 2.81} & 10.36 
\scriptsize{$\pm$ 1.66} & 47.76
\scriptsize{$\pm$ 1.20} & 8.26 \scriptsize{$\pm$ 1.57} & 2.69 \scriptsize{$\pm$ 1.17} & 13.58 \scriptsize{$\pm$ 1.35}\\
SciBERT & 0.69 \scriptsize{$\pm$ 0.29} & 5.65 \scriptsize{$\pm$ 1.36} & 4.60 
\scriptsize{$\pm$ 1.05} & 34.73
\scriptsize{$\pm$ 1.30} & 3.00 \scriptsize{$\pm$ 0.86} & 1.17 \scriptsize{$\pm$ 0.65} & 8.91 \scriptsize{$\pm$ 1.03}\\
\bottomrule
\end{tabular}
\label{tab: 150 candidate}
\end{table}

\section{Model Details}
\label{sec:experiment setup}
\subsection{Retraining Details}
\label{ssec:retrain details}
\begin{figure}[ht]
        \begin{subfigure}[b]{0.5\textwidth}
        \includegraphics[width=\textwidth]{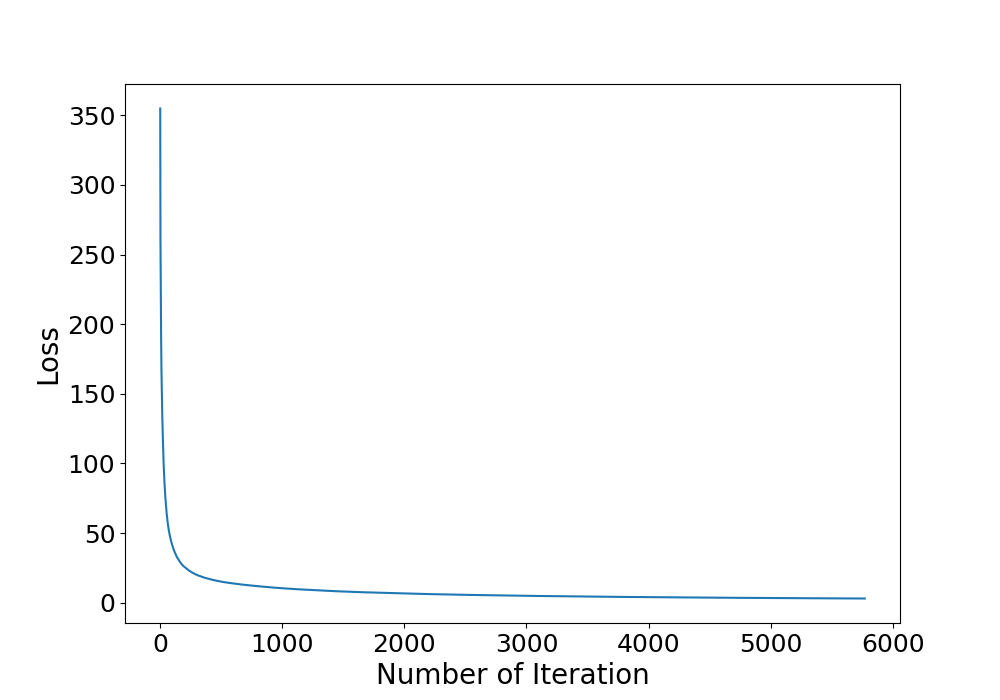}
        \caption{SPECTER}
        \label{fig:training_of_specter}
    \end{subfigure}
    \hfill
    \begin{subfigure}[b]{0.5\textwidth}
        \includegraphics[width=\textwidth]{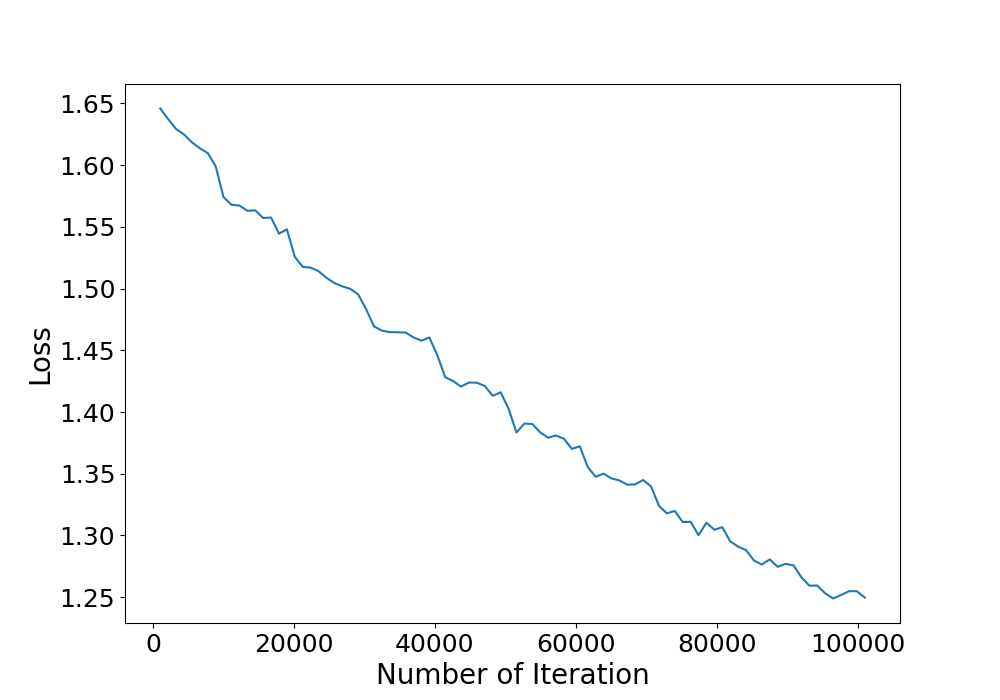}
        \caption{SciBERT}
        \label{fig:training_of_scibert}
    \end{subfigure}
    \caption{Retrained Models}
    \label{fig:retrained_models}
\end{figure}

We retrained SPECTER and SciBERT on our corpus of 360k CS related papers with citation information. In particular, we retrained SPECTER for 4 epochs, but noticed decreasing performance beyond 1 epoch. Therefore, only the epoch 1 checkpoint is kept for SPECTER\textsubscript{ID}. We leveraged the citation signals from Semantic Scholar \cite{semantic_scholar} to train the SPECTER model. Abstract A is related to abstract B if A directly cites B or if A cites a paper that cites B. Our corpus consists of 282,121 abstracts with related abstracts by the above definition. Among these abstracts, each has an average of 76.327 related abstracts. Consequently, we utilize a total of 21,533,311 triplets during the training of SPECTER. We retrained SciBERT for 10 epochs using 8 GPUs in 12 hours. Checkpoint 10 is kept for SciBERT\textsubscript{ID}. Figure \ref{fig:retrained_models} plots the training loss curve for SPECTER and SciBERT.

\subsection{Model Specifications}
\label{ssec: model specifications}

\subsubsection{Sentence-level Embedding Models}

Both ANCE \cite{ance} and SentBERT \cite{sentbert} are trained for sentence-level tasks. Instead of using these models to embed the entire query and abstract, we use them to embed individual sentences of query and abstract.

Approximate nearest neighbor Negative Contrastive Estimation(ANCE) FirstP is a BERT-based model that conducts text retrieval in dense multi-vector representations. The model optimizes the representation of data during training by selecting challenging negative samples using an Approximate Nearest Neighbor (ANN) search. The FirstP model uses the first 512 tokens of the document. The method’s implementation comes from \href{https://huggingface.co/sentence-transformers/msmarco-roberta-base-ance-firstp}{https://huggingface.co/sentence-transformers/msmarco-roberta-base-ance-firstp}. 

Sentence-BERT is an enhancement of the traditional BERT-based transformer model optimized for sentence-level similarity tasks. It uses Siamese and triplet network structures to encode documents into semantically meaningful sentence embeddings. The model is trained on Stanford Natural Language Inference (SNLI) corpus \cite{SNLI} and the Multi-Genre Natural Language Inference (MultiNLI) corpus \cite{MultiNLI}. The method’s implementation comes from \href{https://www.sbert.net/#}{https://www.sbert.net/\#}. 


\subsubsection{Passage Retrieval Models}

ColBERT-v2 \cite{colbertv2}, ERNIE-Search \cite{ernie}, SimLM \cite{simlm}, SPLADE-v2 \cite{spladev2}, and RocketQA-v2 \cite{rocketqa_v2} are models specifically engineered for passage retrieval tasks.

ColBERT-v2 is a retrieval model that incorporates lightweight late interactions on dense multi-vector representations. The BERT-based ColBERT-v2 model is trained on  MS MARCO \cite{msmarco}. By incorporating an aggressive residual compression mechanism and a denoised supervision strategy, ColBERT-v2 achieves more efficient retrieval while maintaining high-quality results. The method’s implementation comes from \href{https://github.com/stanford-futuredata/ColBERT}{https://github.com/stanford-futuredata/ColBERT}. 

ERNIE-Search is a neural retrieval model optimized for open-domain question answering tasks. Built on a pre-trained language model, ERNIE-Search leverages a dual-encoder architecture and introduces a novel method of cross-architecture distillation to improve its performance. The model is trained on MS MARCO and NQ \cite{msmarco, NQ}. The model’s implementation comes from \href{https://huggingface.co/docs/transformers/model_doc/ernie}{https://huggingface.co/docs/transformers/model\_doc/ernie}.

Similarity matching with Language Model pre-training (SimLM), is a retrieval model that embeds queries and abstracts into single-vector representations. It employs a simple bottleneck architecture that encodes documents into a dense vector through self-supervised pre-training. SimLM model is fine-tuned on MS MARCO passage corpus. The method’s implementation comes from \href{https://github.com/microsoft/unilm/tree/master/simlm}{https://github.com/microsoft/unilm/tree/master/simlm}. 

Sparse Lexical and Expansion Model for First Stage Ranking (SPLADE-v2) is a model that offers highly sparse representations for documents and queries for information retrieval tasks. The model is trained on MS MARCO passage ranking dataset. The method’s implementation follows the implementation in BEIR evaluation which comes from \href{https://github.com/beir-cellar/beir/blob/main/beir/retrieval/models/splade.py}{https://github.com/beir-cellar/beir/blob/main/beir/retrieval/models/splade.py}.

RocketQA-v2 is a model that incorporates dynamic list-wise distillation mechanism for jointly training the retriever and the re-ranker. The model is trained on MS MARCO and NQ. The model’s method takes a query and abstracts content as input and outputs their matching scores for the reranking process.
The method’s implementation comes from \href{https://github.com/PaddlePaddle/RocketQA}{https://github.com/PaddlePaddle/RocketQA}.

\subsubsection{Document Similarity Models}

SPECTER \cite{specter} and Aspire \cite{aspire} are models designed for document-level similarities.

Scientific Paper Embeddings using Citation-informed TransformER (SPECTER), is a transformer-based model that is fine-tuned on scientific documents. It generates document-level embedding of scientific documents based on the pretrained SciBERT \cite{scibert} on citation graph. SPECTER was fine-tuned on a subset of the Semantic Scholar corpus. \textcolor{black}{SPECTER 2.0, pre-trained on a collection of newer papers published after 2018, is the successor to SPECTER and is capable of generating task specific embeddings for scientific tasks when paired with adapters. In our experiment, we use the default adaptor.} It should be noted that SPECTER uses L2 distance between vectors instead of cosine-similarity.

The method’s implementation comes from \href{https://github.com/allenai/specter/tree/master}{https://github.com/allenai/specter/tree/master} and \href{https://huggingface.co/allenai/specter2}{https://huggingface.co/allenai/specter2}.

Aspire \cite{aspire} is a document similarity model that flexibly aggregates over fine-grained sentence-level aspect matches. It embeds the query and abstract as multiple vectors. Though each vector represents an individual sentence in the text, it is dependent on its surrounding context. The query and abstract are both embedded once to get the multi-vector representation. The similarity between query and abstract is calculated using the text-supervised single match (TSAspire) method and the optimal-transport multi-match (OTAspire) method, described in \cite{aspire}.
The method's implementation comes from \href{https://github.com/allenai/aspire}{https://github.com/allenai/aspire}.

\subsubsection{General Text Embedding Models}

We also use ada-002 \cite{openaiada}, E5-Large-V2 \cite{e5largev2} and LLAMA \cite{llama} are general text embedding models. These models are not restricted by the size of the context window and can embed longer texts. We only test these models on the full-query DORIS-MAE task \ref{ssec:read the whole query once}.

Ada-002 (ada) is a text embedding model provided by OpenAI. Ada
leverages a transformer-based architecture and incorporates several novel techniques to enhance its embedding quality. The model employs noise reduction techniques to improve the reliability and consistency of the generated embeddings. We use ada-002 via \href{https://platform.openai.com/docs/guides/embeddings}{OpenAI's API}.

E5-Large-V2 300m is a text embedding model developed by Weakly-Supervised Contrastive Pre-training \cite{e5largev2}. The model has demonstrated strong results on benchmarks such as BEIR \cite{beir}. The implementation comes from huggingface at: \href{https://huggingface.co/intfloat/e5-large-v2}{https://huggingface.co/intfloat/e5-large-v2}.

Large Language Model Meta AI (LLAMA), is state-of-the-art foundational LLM released by Meta. The 7 billion model has been trained on about one trillion tokens. We applied for the weights of the 7 billion parameter version. We use the LLAMA to generate the text embedding by extracting the outputs from the penultimate layer of the model.

\section{Author Statement}
\label{sec:author statement}
The authors hereby affirm that we are the sole authors of the submitted manuscript and retain full responsibility for its contents. The authors warrant that this work is original, has not been published elsewhere, and is currently not under consideration for publication elsewhere. In the event of any violation of legal rights, the authors fully accept and bear all the repercussions and consequences. Furthermore, the authors confirm that all data used in this research complies with the necessary licensing requirements.  The authors acknowledge that non-compliance with these statements may lead to the retraction of our work and possible legal consequences.


\clearpage
\section{Datasheet For DORIS-MAE}
\label{sec:Datasheet for DORIS-MAE}

This document is based on \textit{Datasheets for Datasets} by Gebru \textit{et
al.} \cite{gebruDatasheetsDatasets2020}. Please see the most updated version
\href{http://arxiv.org/abs/1803.09010}{here}.

All questions that are not applicable or have a negative answer are omitted for brevity.

 \subsection{ MOTIVATION}

        \textbf{Q: For what purpose was the dataset created?}
    
    
        A: The dataset was created for the purpose of evaluating and improving the performance of scientific document retrieval systems on complex, multi-intent user queries. This work aims to fill the gap in existing resources that often fail to capture the complexity and multifaceted nature of queries typical in a scientific research context.

    \dotfill
    

    \textbf{Q: Who created this dataset (e.g., which team, research group) and on behalf
    of which entity (e.g., company, institution, organization)?}
    
    A: Authors will remain anonymous until the acceptance of our work. 
    
    \dotfill

   \textbf{Q: What support was needed to make this dataset?}

    A: Funders will remain anonymous until the acceptance of our work.

\dotfill

  \subsection{COMPOSITION}

    \textbf{Q: What do the instances that comprise the dataset represent (e.g., documents,
    photos, people, countries)?}

    A: The dataset include several types of data instances.
    \begin{itemize}
        \item Multifaceted complex queries in text, created by human experts.
        \item Hierarchical lists of aspects and sub-aspects for each complex query, created by human experts.
        \item Candidate pools of potentially relevant paper abstracts from arXiv for each complex query.
        \item Annotations (human-made and ChatGPT-made) for each aspect-abstract question pair. 
    \end{itemize}
    \dotfill

\textbf{Q: How many instances are there in total (of each type, if appropriate)?}

A:  It has 50 complex queries containing between 95 to 217 words, 50 Hierarchical lists of aspects and sub-aspects, 50 Candidate pools, 83,591 ChatGPT-made annotations and 250 human-made annotations. 

\dotfill 

    \textbf{Q: Does the dataset contain all possible instances or is it a sample (not
    necessarily random) of instances from a larger set?}

    A: The dataset contains 50 queries in the fields of ML, CV, AI, and NLP. 
    These queries are manually written. As described in Section \ref{ssec:query formation}, queries are formulated by first randomly sampling an abstract from the CS corpus, which contains 360k papers from 2011-2021 posted on arXiv, and designing queries based on these abstracts. Each candidate pool is selected from the CS corpus, which should represent close to the entirety of CS research papers during that time period. 
    
\dotfill

    \textbf{Q: What data does each instance consist of?}

    A: See Section \ref{ssec:doris-mae structure} for a detailed breakdown of each data instance. 
    
    \dotfill

    \textbf{Q: Is there a label or target associated with each instance?}

    A: For each aspect-abstract pair, an annotation score from 0-2 is provided. 
    
    \dotfill
    
    \textbf{Q: Are there recommended data splits (e.g., training, development/validation,
    testing)?}
    

    A: The entire DORIS-MAE dataset consists of 83,591 annotated question pairs of (aspect/sub-aspect and abstract). In particular, the dataset contains a development set of 90 manually annotated pairs for prompt optimization, a test set of 250 manually annotated pairs for hypothesis testing, and 83,591 pairs to compute the relevance rankings of abstracts for each candidate pool. These relevance rankings serve as a test benchmark for evaluating various retrieval models. 
    
    \dotfill
    
    
    \textbf{Q: Are there any errors, sources of noise, or redundancies in the dataset?}

    A: The annotation of the aspect-abstract pair may contain some noise caused by GPU non-determinism. Though we have verified that ChatGPT's performance is on-par with humans', we have documented the possible inconsistencies in Section \ref{ssec:error analysis for chatgpt's annotation}. 
    
\dotfill
    
    \textbf{Q: Is the dataset self-contained, or does it link to or otherwise rely on
    external resources (e.g., websites, tweets, other datasets)?}

    A: The dataset is self-contained.

    \dotfill
    
    

    
    
    
    
    

    

   \subsection{COLLECTION}

    \textbf{Q: How was the data associated with each instance acquired?}

    A: The abstracts in the candidate pools and the corpus are from arXiv\footnote{https://arxiv.org/}. The queries/aspects/sub-aspects are manually created. The annotations are generated by chatgpt-3.5-turbo-0301. 
    
    \dotfill
    
    \textbf{Q: Over what timeframe was the data collected?}

    A: The abstracts in the candidate pools and the corpus are from papers from 2011 to 2021. The queries are created by the authors in 2023. 
    
    \dotfill
    
    \textbf{Q: What mechanisms or procedures were used to collect the data?}

    A: 
    The queries/aspects/sub-aspects are manually created. The abstracts in the candidate pools/corpus are collected from the arXiv API. The citation signals of the papers are collected from Semantic Scholar API\cite{semantic_scholar}. The annotations are generated by chatgpt-3.5-0301 via OpenAI API. The retireval models are retrained using 8 NVIDIA GeForce RTX 2080 Titan GPUs. 
    
    \dotfill
    
    \textbf{Q: What was the resource cost of collecting the data?}

    A: All costs are detailed in Section \ref{ssec:scalability of anno-gpt} and Section \ref{ssec:annotation scalability}. Overall, the API costs less than \$100. 

    \dotfill
    
    \textbf{Q: If the dataset is a sample from a larger set, what was the sampling
    strategy?}
    
    A: The candidate pools are sampled from the corpus by a variety of IR and NIR methods. The queries are created by human experts from existing paper abstracts, which are randomly sampled from the corpus. 
    
    \dotfill
    
    \textbf{Q: Who was involved in the data collection process (e.g., students,
    crowdworkers, contractors) and how were they compensated (e.g., how much
    were crowdworkers paid)?}
    
    A: 
    Only the authors are involved in the data annotation process and compensation is NA. 
    
    \dotfill
    

    
    






    \subsection{USES}

    \textbf{Q: Has the dataset been used for any tasks already?}

    A: It is used to benchmark 16 diverse retrieval models for the DORIS-MAE task in Section \ref{sec:experiment results}. 
    
    \dotfill
    
    \textbf{Q: Is there a repository that links to any or all papers or systems that use the dataset?}
         
    A: \href{https://github.com/Real-Doris-Mae/Doris-Mae-Dataset}{https://github.com/Real-Doris-Mae/Doris-Mae-Dataset} 
    
    \dotfill
    
    \textbf{Q: What (other) tasks could the dataset be used for?}

    A: It could serve as a dataset to evaluate query decomposition models. 

    \dotfill
   
    
    






   \subsection{DISTRIBUTION}

    \textbf{Q: Will the dataset be distributed to third parties outside of the entity
    (e.g., company, institution, organization) on behalf of which the dataset
    was created?}

    A: The dataset is released at Zenodo, and can be accessed \href{https://doi.org/10.5281/zenodo.8035110}{https://doi.org/10.5281/zenodo.8035110}. 

    \dotfill

    \textbf{Q: How will the dataset will be distributed?}

    A: The dataset will be distributed on GitHub and Zenodo. Its DOI is \href{https://doi.org/10.5281/zenodo.8035110}{https://doi.org/10.5281/zenodo.8035110} 

    \dotfill

    \textbf{Q: When will the dataset be distributed?}

    A: It was released on June 13, 2023. 

    \dotfill

    \textbf{Q: Will the dataset be distributed under a copyright or other intellectual
    property (IP) license, and/or under applicable terms of use (ToU)?}

    A: The dataset is distributed under the CC-BY-NC license. For more details, see Section \ref{ssec:dataset license}. 
    


    





\subsection{MAINTENANCE}

    \textbf{Q: Who is supporting/hosting/maintaining the dataset?}

    A: The authors on GitHub and Zenodo. 
    

    \dotfill

    \textbf{Q: How can the owner/curator/manager of the dataset be contacted (e.g., email
    address)?}

 A: One could raise issues on GitHub or contact authors by emails, which will be revealed upon paper's acceptance. 
 

    \dotfill


    
    
    \textbf{Q: Will the dataset be updated (e.g., to correct labeling errors, add new
    instances, delete instances)?}

    A: Yes, the dataset will be updated as needed to correct errors. 

    \dotfill

    \textbf{Q: Will older versions of the dataset continue to be
    supported/hosted/maintained?}

    A: DOIs and downloadable links for older versions of the dataset will be documented on GitHub. 

    \dotfill

    \textbf{Q: If others want to extend/augment/build on/contribute to the dataset, is
    there a mechanism for them to do so?}

    A: Yes. For scalability see Section \ref{ssec:scalability of anno-gpt} and Section \ref{ssec:annotation scalability}. For future work see Section \ref{sec:conclusion and future work}.



\end{document}